\documentclass[12pt]{iopart}

\usepackage{iopams}
\expandafter\let\csname equation*\endcsname\relax
\expandafter\let\csname endequation*\endcsname\relax
\usepackage{amsmath}

\usepackage[utf8]{inputenc}
\usepackage{ragged2e}
\usepackage{lipsum}

\usepackage{calligra}
\usepackage{graphicx,mathtools,amssymb,amsmath,amsthm,amsfonts,epsfig,epsf}
\usepackage[outdir=./]{epstopdf}
\usepackage[dvipsnames]{xcolor}
\usepackage[linktocpage]{hyperref}
\hypersetup{
    colorlinks=true,
    linkcolor=blue,
    filecolor=magenta,      
    urlcolor=BrickRed,
    citecolor=BrickRed,
}
\hypersetup{pdfstartview=}	
\usepackage{tensor}	
\usepackage{csquotes}
\usepackage{mathrsfs}
\usepackage{multirow}

\usepackage{subfig}
\usepackage{graphicx}
\usepackage{amsmath}

\newcommand{\gb}{\mathcal{G}}

\newcommand{\be}{\begin{equation}}
\newcommand{\ee}{\end{equation}}

\begin{document}

\allowdisplaybreaks

\title[Black hole minimum size  and scalar charge  in shift-symmetric theories]{Black hole minimum size  and scalar charge  in shift-symmetric theories}




\author{Farid~Thaalba$^{1,2}$,
Georgios~Antoniou$^{1,2}$,
Thomas~P.~Sotiriou$^{1,2,3}$}

\address{$^1$ Nottingham Centre of Gravity, Nottingham NG7 2RD, United Kingdom}
\address{$^2$ School of Mathematical Sciences, University of Nottingham, University Park, Nottingham NG7 2RD, United Kingdom}
\address{$^3$ School of Physics and Astronomy, University of Nottingham, University Park, Nottingham NG7 2RD, United Kingdom}

\begin{abstract}
It is known that, for shift-symmetric scalars, only a linear coupling with the Gauss-Bonnet (GB) invariant can introduce black hole hair. Such hairy black holes have a minimum mass, determined by the coupling of this interaction, and a scalar charge that is uniquely determined by their mass and spin for a fixed value of that coupling. Here we explore how additional shift-symmetric interactions affect the structure of the black hole, the value of the minimum mass, and the scalar charge.
\end{abstract}

\maketitle

\section{Introduction}
In recent years we have been presented with an unparalleled view of the strong field gravitational regime. The increasing number of gravitational-wave observations from the LIGO-VIRGO-KAGRA collaborations and the black hole images from the EHT collaboration have given us direct access to high-energy gravitational processes, including coalescing compact objects \cite{LIGOScientific:2016aoc,LIGOScientific:2018mvr,LIGOScientific:2020ibl,LIGOScientific:2021usb,LIGOScientific:2021djp}, accretion disks and shadows of supermassive black holes. The direct view of the highly dynamical, strong-field regime that current and future observations offer will give us an unprecedented chance to test the nature of gravity and search for new fundamental fields \cite{Barack:2018yly,Sathyaprakash:2019yqt,Barausse:2020rsu,Kalogera:2021bya,LISA:2022kgy}.

Indeed, black holes in general relativity (GR) are fully characterised by only their mass and spin and by the Kerr metric \cite{Hawking:1971vc,Carter:1971zc}. They can also carry an electric charge in principle, but this is expected to be entirely negligible for astrophysical black holes. Standard Model fields are not expected to endow black holes with any additional parameters \cite{Bekenstein:1972ny,Teitelboim:1972qx}, known as ``hair'', and hence any deviation from this picture in observations could imply the discovery of the new fundamental field or a breakdown of GR. Scalar fields have received particular attention in this context. There exist however several no-hair theorems which dictate that scalars cannot leave an imprint on quiescent black holes \cite{Sotiriou:2015pka,Herdeiro:2015waa}. These theorems cover self-interacting scalars that potentially couple to the Ricci scalar and/or have a noncanonical kinetic term (scalar-tensor theories) \cite{Bekenstein:1995un,Sotiriou:2011dz}. However, more general nonminimal couplings, such as couplings to higher-order curvature invariants, are known to evade them, {\em e.g.}~\cite{Mignemi:1992pm,Kanti:1995vq,Yunes:2011we}.

The case of shift-symmetric scalars is of particular interest when it comes to black hole hair, both observationally and theoretically. Strong gravity observations probe length (or curvature) scales of kilometres. Massive scalar profiles around compact objects are expected to decay exponentially and the characteristic scale for this decay is set by the inverse of the mass. Hence, these observations are expected to probe only ultralight or massless scalars. Shift symmetry, {\em i.e.}~ invariance under $\phi\to \phi+$constant, is the symmetry that protects scalars from acquiring a mass. Consequently, strong gravity observations effectively probe scalars that exhibit either this symmetry or very small violations of it. Interestingly though, shift symmetry implies that the equation of motion of the scalar can be written as a conservation of a current, and this very property was used in \cite{Hui:2012qt} to prove a powerful no-hair theorem. It was subsequently shown in \cite{Sotiriou:2013qea} that a linear coupling between the scalar, $\phi$, and the Gauss-Bonnet invariant, ${\cal G}$, evades this theorem and that this coupling term is unique in this regard. It is worth noting that the linear coupling term $\phi\, {\cal G}$ would arise in a small coupling or small $\phi$ expansion of the exponential coupling $e^\phi {\cal G}$ that had already been known to introduce black hole hair \cite{Mignemi:1992pm,Kanti:1995vq,Yunes:2011we}.

Indeed, hairy black hole solutions in the  $\phi\, {\cal G}$ and in the $e^\phi {\cal G}$ cases share two key properties \cite{Sotiriou:2014pfa}. First, the scalar charge is not an independent parameter, but it is instead fixed with respect to the black hole mass and spin by a regularity condition on the horizon. Second, for any fixed value of the coupling constant that controls these terms ({\em i.e.}~any given theory within the class), black holes have a minimum mass, controlled by the value of that coupling constant. The scope of this paper is to focus on the linear coupling case and determine how these two properties are affected by the presence of additional shift-symmetric (derivative) interactions in the action.

It is worth emphasising that having a minimum mass for black holes in the $\phi\, {\cal G}$ model leads to a strong constraint on the coupling constant of this term, coming from the lightest black hole observed \cite{2022arXiv220710692F, Charmousis:2021npl}. 
Most other observations are sensitive to the scalar charge but this gets converted to a constrain on the same coupling constant using the relation that fixes the scalar charge in terms of the black hole mass (and spin) {\em e.g.~}\cite{Lyu:2022gdr,Perkins:2021mhb,Maselli:2020zgv,Maselli:2021men}. In the case of extreme mass ratio inspirals (EMRIs), a scaling of the scalar charge with respect to the black hole mass inspired by the $\phi\, {\cal G}$ theory proved to be a crucial ingredient to drastically simplify the modelling for a vastly broader class of nonminimally coupled scalar \cite{Maselli:2020zgv}. Note that, although there are good reasons to believe that such terms can remain subdominant when modelling binary dynamics and gravitational wave radiation \cite{Witek:2018dmd,Maselli:2020zgv}, they could still have a crucial effect on the properties of the sources, including their quasi-normal ringing \cite{Hui:2021cpm}, and their dependence on the coupling constants of the theory \cite{Saravani:2019xwx}. This can  affect how observations get translated to bounds for these coupling constants. 
 Considering also that Effective field theory (EFT) strongly suggests that additional terms should be present, it is rather pertinent to understand their contributions to relations between mass and charge and to check whether they affect the minimum size of black holes. 
 
The first step in this direction has already been made in \cite{Saravani:2019xwx}. It was shown there that, for the most general shift symmetric action that leads to second-order equations upon variation (shift-symmetric Horndeski theory) and respects local Lorentz symmetry, the scalar charge $Q$ for black holes is given by $4\pi Q= \alpha \int_{\cal H} n_a {\cal G}^a$, where ${\cal H}$ denotes the Killing horizon, $n^a$ its normal,  $\mathcal{G} ^a$ in implicitly defined as ${\cal G}=\nabla_a {\cal G}^a$, and $\alpha$ is the coupling constant associated with the $\phi {\cal G}$ term. As expected from the discussion above, the charge vanishes if the $\phi {\cal G}$ term is absent. The value of ${\cal G}^a$ does however depend on any additional couplings, with corrections with respect to the value ${\cal G}^a$ would have in GR suppressed by the mass scales that correspond to these couplings. If one assumes continuity as these couplings are driven to zero and that their characteristic energy scales are similar to that of $\alpha$ ({\em i.e.}~no hierarchy of scales), then one expects corrections to the charge and its scaling with the mass of the black hole to be subdominant. How much so is a matter of further exploration. Moreover, this expression for the charge does not give any information on the minimum size of black holes.

 The structure of the rest of the paper is as follows: in Sec.~\ref{sec:classification}, we will review the necessary theoretical background and present the class of theories we will consider.  In Sec.~\ref{sec:spherical}, we will lay out the problem in a static, spherically symmetric setup and also  consider the behaviour of a scalar field in a Schwarzschild background (decoupling) as a warm-up. In Sec.~\ref{sec:perturbative} we will derive hairy black hole solutions working perturbatively in the coupling $\alpha$ and analyse their properties, while in Sec.~\ref{sec:numerical}, we will present numerical results and a comparison with the perturbative ones. Finally, in Sec.~\ref{sec:conclusions}, we present our conclusions.
 
\section{Theoretical background}
\label{sec:classification}

\subsection{Shift-symmetric Horndeski gravity and black hole hair}
Horndeski's theory is the most general four-dimensional diffeomorphism-invariant theory involving a metric tensor and a scalar field that leads to second-order field equations upon variation \cite{Horndeski:1974wa,Deffayet:2009mn}. We will restrict ourselves to shift symmetric theories. The shift-symmetric Horndeski action is then \cite{Sotiriou:2014pfa}
\begin{equation}
\label{Horndeski}
S=\frac{1}{2k}\sum_{i=2}^{5}\int \mathrm{d}^4x\,\sqrt{-g}\mathcal{L}_i+S_{M},
\end{equation}
where each sub-Lagrangian $\mathcal{L}_i$ is given by
\begin{align}
\label{L2}
\mathcal{L}_2 = & \, G_2(X),\\
\label{L3}
\mathcal{L}_3 = & \, -G_3(X)\Box\phi,\\
\label{L4}
\mathcal{L}_4 = & \, G_4(X)\mathcal{R}+G_{4X}[(\Box\phi)^2-(\nabla_\mu\nabla_\nu\phi)^2],\\
\notag
\mathcal{L}_5 = & \, G_5(X)G_{\mu\nu}\nabla^\mu\nabla^\nu\phi
- \frac{G_{5X}}{6}\left[\left(\Box\phi\right)^3 -3\Box\phi(\nabla_\mu\nabla_\nu\phi)^2+2(\nabla_\mu\nabla_\nu\phi)^3\right],
\end{align}
where we have defined $X=-\nabla_\mu\phi\nabla^\mu\phi/2$, $(\nabla_\mu\nabla_\nu\phi)^2=\nabla_\mu\nabla_\nu\phi\nabla^\mu\nabla^\nu\phi$, $(\nabla_\mu\nabla_\nu\phi)^3= \nabla_\mu\nabla_\nu\phi\nabla^\nu\nabla^\lambda\phi\nabla_\lambda\nabla^\mu\phi$, $G_{iX}=\partial G_{i}/\partial X$, $\mathcal{R}$ is the Ricci scalar, and $G_{\mu\nu}$ is the Einstein tensor. We have also defined $k=8\pi G/c^4$ with $S_M$ being the matter action. Matter is assumed to only couple minimally to the metric, that is, we are working in the so-called Jordan frame. 

Shift-symmetry implies that the field equation for the scalar can be written as a conservation of a current,
\begin{equation}
\label{Jcons}
\nabla_\mu J^\mu = 0
\end{equation}
The current is given by 
\begin{equation}
    \begin{split}
        J^\mu =\,& -\partial^\mu\phi \bigg( G_{2X} - G_{3X} \Box \phi +G_{4X} \mathcal{R}
        + G_{4XX} \left[ (\Box \phi)^2 -(\nabla_\rho\nabla_\sigma\phi)^2 \right]\\
        &+G_{5X}G^{\rho\sigma}\nabla_{\rho}\nabla_{\sigma}\phi-\frac{G_{5XX}}{6} \left[ (\Box \phi)^3 - 3\Box \phi(\nabla_\rho\nabla_\sigma\phi)^2 + 2(\nabla_\rho\nabla_\sigma\phi)^3 \right] \bigg)\\
        &-\partial^\nu X \bigg( - \delta^\mu_\nu G_{3X} + 2 G_{4XX} (\Box \phi \delta^\mu_\nu-\nabla^\mu\nabla_\nu \phi)+G_{5X} G^\mu{}_\nu\\
        &-\frac12 G_{5XX} \big[ \delta^{\mu}_{\nu}(\Box\phi)^2- \delta^{\mu}_{\nu}(\nabla_\rho\nabla_\sigma\phi)^2 -2\Box\phi \nabla^\mu\nabla_\nu\phi+2\nabla^\mu \nabla_\rho \phi \nabla^\rho \nabla_\nu \phi \big]   \bigg)\\
        &+2G_{4X} \mathcal{R}^{\mu}{}_{\rho} \nabla^\rho \phi + G_{5X} \bigg( -\Box \phi \mathcal{R}^\mu{}_\rho \nabla^\rho\phi + \mathcal{R}_{\rho\nu}{}^{\sigma\mu} \nabla^\rho\nabla_\sigma\phi \nabla^\nu\phi\\
        &+\mathcal{R}_\rho{}^\sigma \nabla^\rho\phi  \nabla^\mu\nabla_\sigma\phi \bigg).
    \end{split}
\end{equation}
As discussed in the Introduction, this current conservation equation was used to prove a no-hair theorem in~\cite{Hui:2012qt}. 
It was first shown that, for vacuum, static, spherically-symmetric, asymptotically flat black holes, the only non-vanishing component of the current is the radial one $J^{r}$. It was then argued that $J^{r}$ must  vanish at the horizon, otherwise  $\left(J^{r}\right)^2/g^{rr}$ would diverge there.  Current conservation then implied that $J^{r}$ must be zero everywhere. Finally, it was argued  that $J^{r}=0$ everywhere implies that the scalar field must be constant everywhere. This no-hair theorem generalized to slowly rotating black hole straightforwardly \cite{Sotiriou:2014pfa}.

It was however shown in~\cite{Sotiriou:2013qea} that a linear coupling between $\phi$ and the Gauss-Bonnet invariant, $\gb = \mathcal{R}^{\mu \nu \rho \sigma} \mathcal{R}_{\mu \nu \rho \sigma}-4 \mathcal{R}^{\mu \nu} \mathcal{R}_{\mu \nu}+\mathcal{R}^{2}$,  circumvents this no-hair theorem. Indeed, ${\cal G}$ is a total divergence and so $\phi {\cal G}$ respects shift symmetry. However, consider the action
\begin{equation}
\label{eq:lgbaction}
     S= \frac{1}{2 k}\int \mathrm{d}^4 x\,\sqrt{-g}\bigg[\frac{\mathcal{R}}{2}+X+\alpha\,\phi\,\mathcal{G}\bigg]. 
\end{equation}
The corresponding scalar equation is 
\begin{equation}
\label{eq:lgbeq}
    \Box\phi =-\alpha {\cal G}.
\end{equation}
It can be written as a conservation of a current in the form $\nabla_\mu (\nabla^\mu\phi+\alpha {\cal G}^\mu)=0$, exploiting the fact that ${\cal G}$ is a total divergence. Although it is not obvious, the $\phi {\cal G}$ is part of action \eqref{Horndeski} \cite{Kobayashi:2011nu}. However, it does not admit any constant $\phi$ solutions unless ${\cal G}=0$, which is not the case for black holes and hence they will have to have hair. The apparent contradiction with the no-hair theorem of~\cite{Hui:2012qt} is resolved by the final argument of~\cite{Hui:2012qt} --- that vanishing current implies constant $\phi$ --- relies on the assumption that every single term in the current depends on the gradient of $\phi$. The contribution of the linear coupling to ${\cal G}$ clearly violates this assumption and is indeed unique in this respect \cite{Sotiriou:2013qea}. Interestingly, hairy black hole solutions in this theory violate another assumption of the theory of ~\cite{Hui:2012qt}:  $\left(J^{r}\right)^2/g^{rr}$ diverges on the horizon \cite{Babichev:2017guv}. It was subsequently shown in \cite{Creminelli:2020lxn} that this quantity is not an invariant when  it received a contribution from a linear coupling with ${\cal G}$ and hence there is no reason to impose that it is finite in this case.

In~\cite{Saravani:2019xwx} theories described by action \eqref{Horndeski} where classified as follows: 
\begin{align}
    \textbf{Class-1: } &\mathcal{E}_\phi[\phi=0,\,g]=0,\qquad \forall g,\\[2mm]
    \textbf{Class-2: } &\lim_{g\rightarrow \eta}\mathcal{E}_\phi [\phi=0,\, g]=0.\\[0mm]
    \textbf{Class-3: } &\text{All the rest}.
\end{align}
Class-1 theories are defined by having $\phi=0$ as a solution for any general background; hence, they admit all possible GR solutions. Class-2 theories allow for $\phi=0$ to be realized only for flat spacetimes.  The third class is defined as the complement of the other two. Therefore, class-3 theories admit a non-trivial scalar configuration in flat spacetime as a solution, or flat spacetime is not a solution, and hence they violate Local Lorentz symmetry.  

 At first sight, it appears that classes 1 and 2 are unrelated. On the contrary, it was shown in~\cite{Saravani:2019xwx} that a class-2 Lagrangian can always be expressed as a class-1 Lagrangian plus a contribution from the Gauss-Bonnet invariant, namely:
\begin{equation}
    \mathcal{L}_{(2)}=\mathcal{L}_{(1)}+\alpha\phi\mathcal{G}.
\end{equation}
Consequently, all shift-symmetric non-Lorentz violating Horndeski theories admit all GR solutions, provided that a linear coupling between the scalar and the Gauss-Bonnet invariant is not present. Using this result, it was then shown that the scalar charge $Q$ of a stationary black hole in any theory in classes 1 and 2 is given by
\begin{equation}
\label{eq:charge}
4\pi Q= \alpha \int_{\cal H} n_a {\cal G}^a,
\end{equation}
where ${\cal H}$ denotes the Killing horizon and $n^a$ its normal, as already mentioned in the Introduction. 

\subsection{Our model}

As already mentioned in the Introduction, the hairy black holes of the theory in action \eqref{eq:lgbaction} have two key properties: their scalar charge $Q$ is not an independent parameter, but it is instead determined by their mass (and spin), in accordance with eq.~\eqref{eq:charge}, and they have a minimum mass. In the next Section, we will see in detail how these properties relate to regularity conditions for static, spherically symmetric black holes. Our broader goal is to understand how adding additional shift symmetric terms to action \eqref{eq:lgbaction} would affect these properties. 

To make the calculations more tractable, we will not consider action \eqref{Horndeski}. We will instead restrict ourselves to the following theory 
\begin{equation}
\label{eq:action}
\begin{split}
    S=\frac{1}{2k}\int \mathrm{d}^4 x\,\sqrt{-g}\bigg[&\frac{\mathcal{R}}{2}+X+\alpha\,\phi\,\mathcal{G}
    +\gamma \,G_{\mu\nu}\nabla^\mu\phi\nabla^\nu\phi
    +\sigma X\Box\phi +\kappa\, X^2 \bigg]. 
\end{split}
\end{equation}
This action can be obtained from action \eqref{Horndeski} by selecting  \footnote{Up to a total derivative the term $X\mathcal{R}+(\Box\phi)^2-(\nabla_\mu\nabla_\nu\phi)^2$ is equivalent to $\,G_{\mu\nu}\nabla^\mu\phi\nabla^\nu\phi$. Since, $\int \mathrm{d}^4x \sqrt{-g}\,(\Box\phi)^2=\int \mathrm{d}^4x \sqrt{-g}\,\big[(\nabla_\mu\nabla_\nu\phi)^2+\mathcal{R}_{\mu\nu}\nabla^\mu\phi\nabla^\nu\phi\big]+\int \text{total derivative}\;.$} 
\begin{equation}
    \begin{split}
    G_2(X) & \coloneqq X+\kappa X^2, \\
    G_3(X) & \coloneqq -\sigma X, \\
    G_4(X) &  \coloneqq 1/2 + \gamma X, \\ 
    G_5(X) & \coloneqq -4\alpha\ln|X|.
    \end{split}
\end{equation}
In units where $G=c=1$ the scalar field is  dimensionless while $\alpha, \gamma, \sigma, \kappa$ have dimensions of length squared.
\section{Spherically symmetric setup}
\label{sec:spherical}
In this section, we consider a static and spherically symmetric background, described by the following metric,
\begin{equation}
\label{eq:metric}
    \mathrm{d}s^2=-A(r)\mathrm{d}t^2+\frac{1}{B(r)}\mathrm{d}r^2+r^2 \mathrm{d}\Omega^2,
\end{equation}
while the scalar field depends only on the radial coordinate, $\phi=\phi(r)$. 

\subsection{Shift-symmetric current}

The only non-vanishing component of the current $J^\mu$ is $J^r$, given by
\begin{align}
\label{eq:spherical_current}
\begin{split}
    J^r=& 
    -B\phi'(1-\kappa B\phi^{\prime 2}) -\sigma \phi^{\prime 2} \frac{r A' + 4 A}{2rA}B^2
    +\gamma \phi' \frac{2A B -2A + 2rA'B}{r^2A}B\\[2MM]
    &
    + \alpha \frac{4(1-B) B A'}{r^2 A}\,,
\end{split}
\end{align}\\
where a prime denotes a derivative w.r.t. the radial coordinate $r$. As discussed earlier, and according to the classification of \cite{Saravani:2019xwx}, the current can be separated into a part for which every term contains $\phi'$ and a contribution by the coupling with ${\cal G}$, as follows
\begin{equation}
    J^r=\tilde{J}^r-\alpha\mathcal{G}^r,\qquad \mathcal{G}^r=\frac{4(B-1) B A'}{r^2 A},
    \label{eq:current}
\end{equation}
The conservation of the current can be straightforwardly integrated:
\begin{equation}
\label{eq:current_2}
    \nabla_\mu J^\mu=0\Rightarrow J^r=\frac{c}{r^2}\sqrt{\frac{B}{A}}. 
\end{equation}
Using \eqref{eq:spherical_current}, one can then determine $\phi'$ and then integrate once more to obtain $\phi(r)$. Due to shift symmetry, $c$ in \eqref{eq:current_2}, is the only meaningful integration constant. Hence, considering also the mass parameter of the black hole, one would have a two-parameter family of solutions. The scalar charge would then be independent. 

However, $\phi'$ evaluated on the horizon of a black hole, $r=r_h$, denoted as $\phi'_h$, generically diverges. 
If we assume that the scalar is regular on the horizon, and hence $\phi'_h$ is finite, and we take into account that at $r=r_h$ we have $A(r_h),B(r_h)\rightarrow0$, then  \eqref{eq:spherical_current} implies that $\tilde{J}^r(r_h)=0$. Evaluating \eqref{eq:current} on the horizon then fixes the value of $c$. As a consequence, the scalar charge ceases to be an independent parameter. After substituting $c$ back in \eqref{eq:current}, and solving with respect to $\tilde{J}^r$ we find:
\begin{equation}
    \tilde{J}^r=\frac{4 \alpha}{r^2}\sqrt{\frac{B}{A}}\bigg[\lim_{r\rightarrow r_h}\bigg(\sqrt{A'B'}\text{ sgn}(B')\bigg)+\left(B-1\right)\sqrt{\frac{B}{A}}A'\bigg]. 
\end{equation}
This is the equation we will be using in the following subsection.

\subsection{Decoupling limit}
As a warm-up,  we consider the scalar in a fixed Schwarzschild background. We look for  solutions that are regular at the horizon and approach a constant value asymptotically. For simplicity, we fix that value to zero, as the value of the constant is irrelevant due to shift symmetry. The $\gamma$ term is not expected to contribute anything to the decoupled equations since it multiplies the Einstein tensor which vanishes at decoupling. The scalar equation on a GR-Schwarzschild background is:
\begin{equation}
    \label{eq:Schwarzschild_BG}
    m r^3 \kappa (2 m-r) \phi'^3+m r^2 \sigma  (2 r-3 m) \phi'^2+m r^4 \phi'+2 \alpha  \left(4 m^2+2 m r+r^2\right)=0 \, .
\end{equation}

First, let us note that if only the $\alpha$-term is present, one can find an analytic solution for the scalar field \cite{Sotiriou:2013qea}:
\begin{equation}\label{eq:sc_GR}
    \phi_\alpha=\frac{2 \alpha  \left(4 m^2+3 m r+3 r^2\right)}{3 m r^3}\, .
\end{equation}
We note that, in this case, no specific restriction on the choices of $\alpha$ is suggested. We then consider the $\sigma$-term in addition to the $\alpha$ one, and we solve for the derivative of the scalar field:
\begin{equation}\label{eq:sc_sigma_GR}
    \phi'_{\alpha\sigma}=\frac{1}{6 m r \sigma -4 r^2 \sigma }\left[r^3-\sqrt{r^6 + 8\alpha\sigma \left(12m^2-2m r - r^2 -\frac{2r^3}{m}\right)}\;\right],\, \alpha\sigma<\frac{2m^4}{3}.
\end{equation}
The quantity under the square root needs to be non-zero and positive, therefore, an existence condition emerges. It is straightforward to see that
\begin{equation}
    \lim_{\sigma\rightarrow 0}\phi'_{\alpha\sigma}=\,\phi'_\alpha\quad , \quad\phi'_{\alpha\sigma}(r\gg r_h)=\phi'_{\alpha}(r\gg r_h)\approx-\frac{2\alpha}{m r^2}\;.
\end{equation}
The inequality condition appearing in \eqref{eq:sc_sigma_GR} imposes an upper bound on the product $\alpha\sigma$. This, in turn, yields an upper bound on $\alpha$ when $\sigma>0$, and a lower one when $\sigma<0$. It is also possible to employ a near-horizon expansion, i.e. $r=r_{h}+\epsilon$, for the two cases discussed above. This yields 
\begin{equation}
    \phi'_{\alpha} = -\frac{3\alpha}{2 m^3} + \mathcal{O}(r-2m) \quad , \quad
    \phi'_{\alpha\sigma} = -\frac{2 m^2 - \sqrt{4 m^4-6 \alpha  \sigma }}{m \sigma } + \mathcal{O}(r-2m)\, ,
\end{equation}
When $\alpha$ is positive, from the near-horizon expressions we can deduce that for $\sigma>0$ ($\sigma<0$) the scalar field fall-off is larger (smaller) than in the $\sigma=0,\,\alpha\ne 0$ case, while in the limit $\sigma\rightarrow -\infty$ we retrieve the trivial solution $\phi_{\alpha\sigma}=0$ for the near-horizon expansion. When $\alpha$ is negative the aforementioned properties are reversed.

The case where $\kappa\ne 0$ is more subtle. By examining the $\phi'^3$ coefficient in \eqref{eq:Schwarzschild_BG}, we see that when the $\kappa$-term is present, the derivative of the scalar at the horizon does not depend on $\kappa$, therefore, we deduce that $\kappa$ does not enter an existence condition analogous to \eqref{eq:sc_sigma_GR}. If one attempts to solve the equation in the region $[r_h+\epsilon,\infty)$, for $\epsilon \ll 1$, it turns out that a regular solution can be found $\forall\,\kappa\in\mathbb{R}$. However, not all of those solutions have the desired asymptotic behaviour, and for large positive values of $\kappa$,  $\phi'_{\alpha\kappa}(\infty)\ne 0$.

\subsection{Existence conditions}
In the previous subsection, we saw how the existence conditions for the scalar equation are affected by the extra terms in the decoupling limit. Here, we derive the existence conditions for black holes beyond decoupling, for the full system of equations. To do so, we assume the existence of a horizon located at $r=r_h$, so that $A_{h^\pm}\rightarrow 0^\pm$, where the $+$ sign corresponds to approaching the horizon from the outside, while the $-$ to approaching it from the inside. By employing near-horizon expansions we obtain the following expression for the second derivative of the scalar at the horizon:
\begin{align}\label{eq:phidd}
    \phi''=&-\frac{\left(4 \alpha  \phi'+r\right) \left\{24 \alpha +\phi'^2 \left[24 \alpha  \gamma+r^2 (8 \alpha +\sigma )\right]+2 r^3 \phi'\right\}}{2 \left\{r^4-96 \alpha ^2+\phi' \left[r^3 (4 \alpha +\sigma )+24 \alpha \gamma r \right]\right\}}\left(\frac{A'}{A}\right)+\mathcal{O}(1).
\end{align}
We note that in order to get a black hole solution with a nontrivial scalar field   it is required that $\left(4 \alpha  \phi'+r\right)\ne 0$ \cite{Sotiriou:2014pfa} (cf. \cite{Kanti:1995vq, Antoniou:2017acq}). Since $(A'/A)_h$ diverges, in order for $\phi$ to be regular at the horizon, it is required that 
$$[24 \alpha +\phi'^2 \left[24 \alpha  \gamma +r^2 (8 \alpha +\sigma )\right]+2 r^3 \phi']_{r_h}=0,$$ therefore, 
\begin{equation}\label{eq:phidh}
    \phi_h'=\frac{\sqrt{r_h^6-576 \alpha ^2 \gamma-24 \alpha  r_h^2 (8 \alpha +\sigma )}-r_h^3}{24 \alpha  \gamma +r_h^2 (8 \alpha +\sigma )}.
\end{equation}
It is possible to derive from equations \eqref{eq:phidd}, and \eqref{eq:phidh} two conditions:  
\begin{align}
    \label{eq:condition_1}
    \textbf{I:}\quad & r_h^6-576 \alpha ^2 \gamma -24 \alpha  r_h^2 (8 \alpha +\sigma ) \ge 0\,,\\
    \label{eq:condition_2}
    \begin{split}
        \textbf{II:}\quad & \left[24 \alpha  \gamma  r_h+(4 \alpha +\sigma ) r_h^3\right] \sqrt{r_h^6-576 \alpha ^2 \gamma -24 \alpha  (8 \alpha +\sigma ) r_h^2}\\
        &+4\alpha\left[r_h^6-576 \alpha ^2 \gamma-24 \alpha  (8 \alpha +\sigma ) r_h^2 \right]\ne 0\,.
    \end{split}
\end{align}
Condition \textbf{I} comes from the requirement that the quantity under the square root in equation \eqref{eq:phidh} needs to be positive. Condition \textbf{II} comes from requiring that the denominator of the fraction in the right-hand side of equation \eqref{eq:phidd} does not vanish on the horizon (where $\phi'$ is given by equation \eqref{eq:phidh}). Although not obvious, these two conditions also guarantee that the denominator of the fraction on the right-hand side of equation \eqref{eq:phidh} does not vanish.

It is worth pointing out that when we consider only the GB term, conditions I and II reduce to the existence condition appearing in \cite{Sotiriou:2014pfa}. Thus, we see that in the non-perturbative approach $\sigma$ but also $\gamma$ enter the existence conditions. In particular, in the case $\gamma=0$, the parameter $\alpha$ has both an upper and a lower bound for either sign of $\sigma$. This will become more clear in sec.~\ref{sec:numerical} where particular choices of the couplings are examined.

\section{Perturbative treatment}
\label{sec:perturbative}

First, we will employ a perturbative approach with respect to the coupling constant  $\alpha$, which is associated with the term that sources the hair. To do that we define the dimensionless parameter $\tilde{\alpha}\equiv \alpha/r_h^2\ll 1$, where the horizon radius $r_h$ is the length scale we associate with our solution. In a similar manner we can define $\tilde{\gamma}=\gamma/r_h^2,\, \tilde{\sigma}=\sigma/r_h^2$ and $\tilde{\kappa}=\kappa/r_h^2$.  For nonzero, small values of $\tilde{\alpha}$ we expect to acquire perturbative deformations to the Schwarzschild solution. Those are expressed through the following expansions for the metric elements
\begin{align}
    A(r)=&\left(1-\frac{2M}{r}\right)\left(1+\sum_{n=1}^{\infty}A_n(r)\tilde{\alpha}^n\right)^2,\label{eq:A_exp} \\
    B(r)=&\left(1-\frac{2M}{r}\right)\left(1+\sum_{n=1}^{\infty}B_n(r)\tilde{\alpha}^n\right)^{-2}, \label{eq:B_exp}\\
    \phi(r)=&\;\phi_0+\sum_{n=1}^\infty \phi_n \tilde{\alpha}^n.
\end{align}
For $\tilde{\alpha}=0$, we retrieve GR minimally coupled to a scalar field, which for the spherically symmetric static configurations yields the Schwarzschild solution. These expansions are substituted in the equations of motion, which are then solved order by order for the unknown coefficients $\{A_n,\,B_n,\,\phi_n\}$. We work out the calculations up to the fifth order in the perturbative parameter $\mathcal{O}(\tilde{\alpha}^5)$. The solutions become very lengthy beyond $2^{\text{nd}}$-order and, therefore, are omitted. However, the expressions for the scalar charge and the mass of the black hole can be written in the following compact form:
\begin{align}
    Q=&\,Q_1\,\tilde{\alpha}+Q_3\,\tilde{\alpha}^3+Q_4(\sigma)\,\tilde{\alpha}^4+Q_5(\gamma,\sigma,\kappa)\,\tilde{\alpha}^5,\label{eq:Qa}\\[4mm]
    M=&\,m+M_2\,\tilde{\alpha}^2+M_3(\sigma)\tilde{\alpha}^3+M_4(\gamma,\sigma,\kappa)\,\tilde{\alpha}^4
    +M_5(\gamma,\sigma,\kappa)\,\tilde{\alpha}^5\label{eq:Ma},
\end{align}
where the coefficients $Q_n$ and $M_n$ can be found in the appendix. Notice that we have to expand to the $5^{\text{th}}$ order in $\tilde{\alpha}$ before we see $\kappa$ contributing to the scalar charge.

The perturbative treatment will break down at some radius. To trace when that happens we simultaneously scan the following expressions
\begin{align}
    A(r)=\,& \bar{A}_0(r)+\bar{A}_2(r)\,{\tilde{\alpha}}^2+\bar{A}_3(r,\sigma)\,{\tilde{\alpha}}^3\, ,\label{eq:A_pert}\\
    B(r)=\,& \bar{B}_0(r)+\bar{B}_2(r,\gamma)\,{\tilde{\alpha}}^2+\bar{B}_3(r,\gamma,\sigma)\,{\tilde{\alpha}}^3\, ,\label{eq:B_pert}\\
    \phi(r)=\,& \phi_0+\phi_1(r)\,{\tilde{\alpha}}+\phi_2(r, \sigma)\,{\tilde{\alpha}}^2+\phi_3(r,\gamma,\sigma,\kappa)\,{\tilde{\alpha}}^3\, ,\label{eq:phi_pert}\\
    \gb(r)=\,& \gb_0(r)+\gb_2(r,\gamma)\,{\tilde{\alpha}}^2+\gb_3(r,\gamma,\sigma)\,{\tilde{\alpha}}^3\, ,\label{eq:GB_pert}
\end{align}
for perturbative inconsistencies. Note that the quantities $\bar{A}_n, \bar{B}_n$ appearing in equations \eqref{eq:A_pert} and \eqref{eq:B_pert} differ from $A_n, B_n$ appearing in equations \eqref{eq:A_exp} and \eqref{eq:B_exp}.
If at some radius $r_{\text{np}}$ terms of different orders of $\tilde{\alpha}$ become comparable in size, the perturbative treatment can no longer be trusted. The coefficients $\phi_n,\,\gb_n$ are given in \ref{sec:appendix_perturbative}. From \eqref{eq:A_pert}-\eqref{eq:GB_pert} we see that even at second-order in $\tilde{\alpha}$, terms involving $\gamma$ appear. We note that in the case where  $\tilde{\gamma},\,\tilde{\sigma},\,\tilde{\kappa}=0$ it was shown in \cite{Sotiriou:2014pfa} that loss of perturbativity occurred at roughly the same radius at which the non-perturbative solutions exhibited a finite area singularity. We will return to this issue in the next section.

\begin{figure}[t]
    \centering
    \includegraphics[width=.47\linewidth]{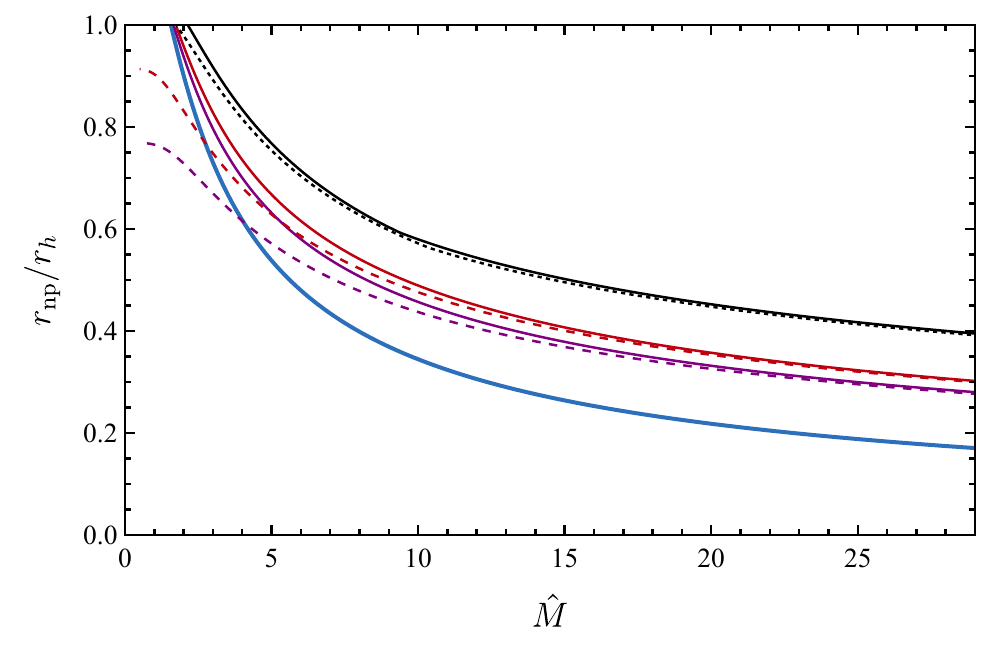}
    \hspace{5mm}
    \includegraphics[width=.47\linewidth]{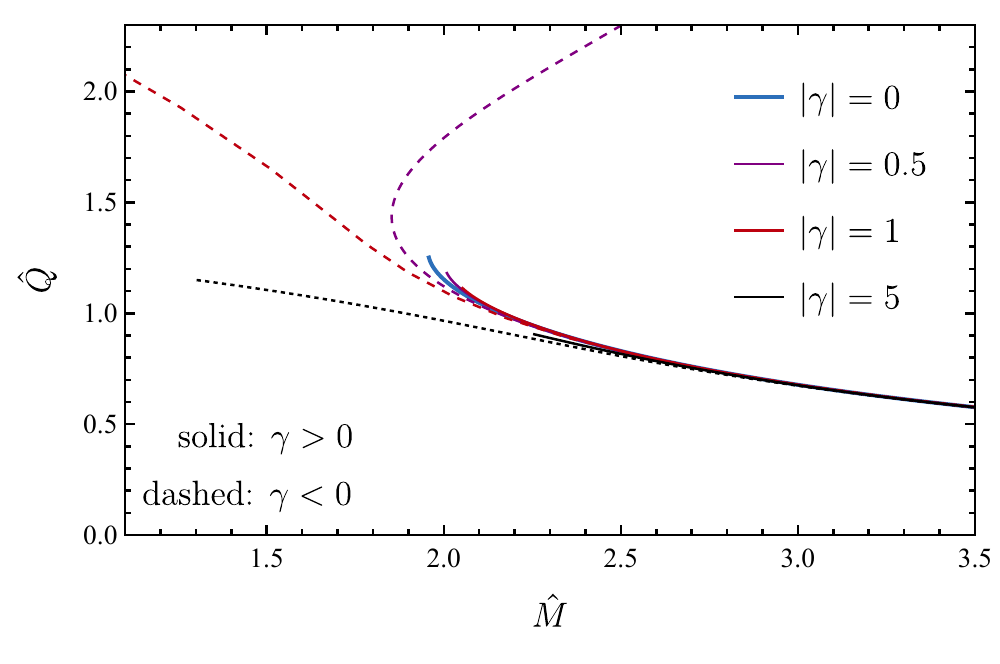}\\
    
    \includegraphics[width=.47\linewidth]{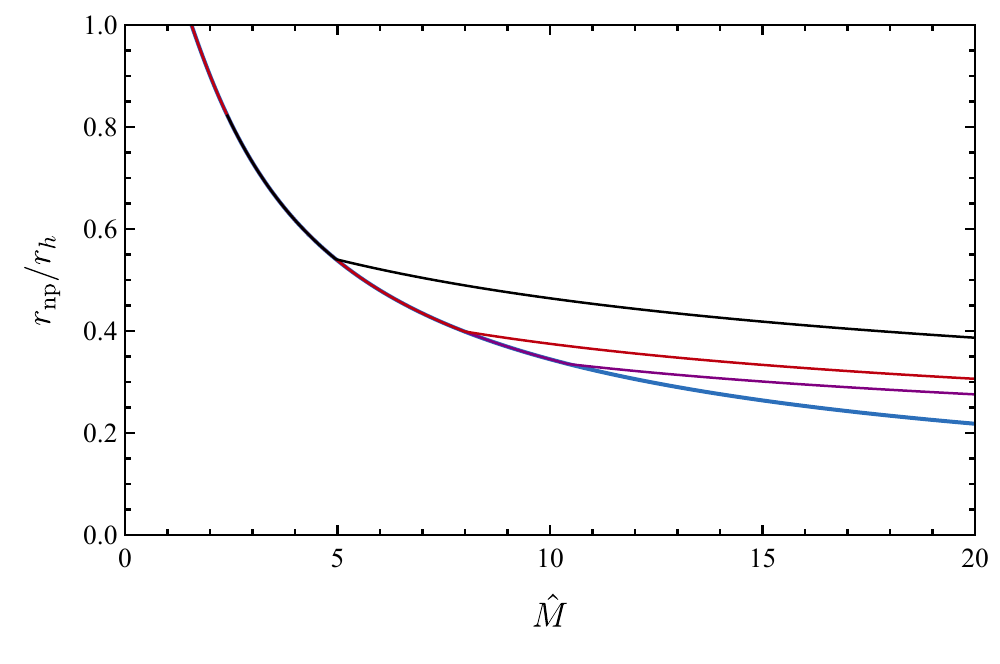}
    \hspace{5mm}
    \includegraphics[width=.47\linewidth]{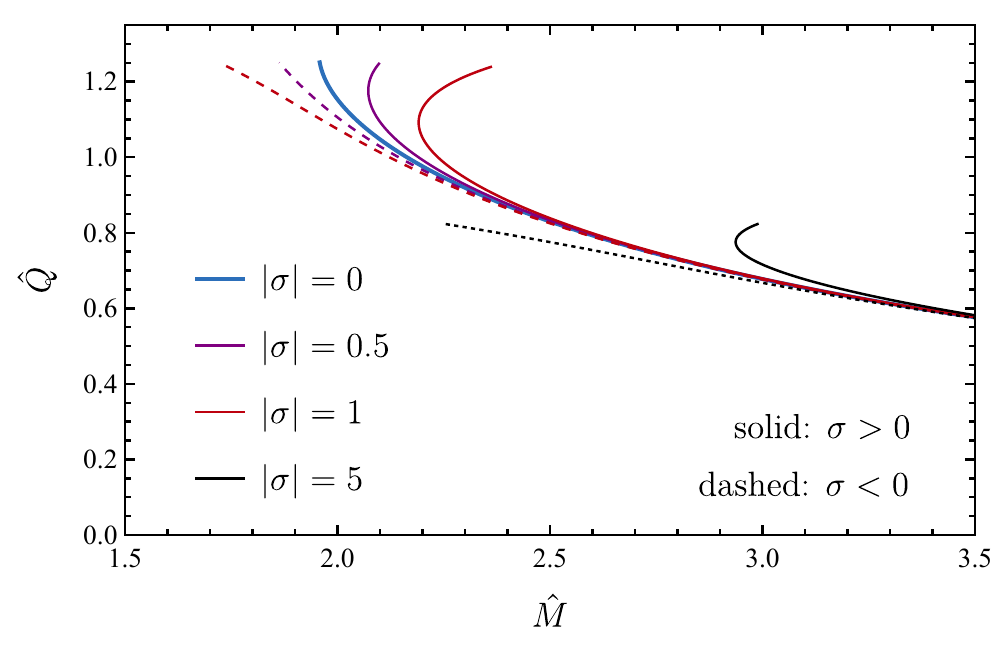}
    
    \caption{\textit{Top left}: Singular radius derived from the perturbative analysis, for $\tilde{\sigma}=0$ and $\tilde{\gamma}=0,\,\pm 0.5,\,\pm 1,\, \pm 5$. \textit{Top right}: Normalized scalar charge and mass derived from the perturbative analysis. \textit{Bottom left}: Singular radius derived from the perturbative analysis, for $\tilde{\gamma}=0$ and $\tilde{\sigma}=0,\,\pm 0.5,\,\pm 1,\, \pm 5$.\textit{Bottom right}: Normalized scalar charge and mass derived from the perturbative analysis.}
    \label{fig:perturbative_treatment}
    
\end{figure}
In the top-left panel of Fig.~\ref{fig:perturbative_treatment} we present the radius $r_{\text{np}}$ denoting the point where the perturbative analysis breaks down, for $\tilde{\gamma}=0,\,\pm 0.5,\,\pm 1,\, \pm 5$ and $\tilde{\sigma}=0$. In the top-right panel of Fig.~\ref{fig:perturbative_treatment} we present the $\hat{Q}$-$\hat{M}$ solution-existence curves for the same choices of $\tilde{\gamma}$. The quantities $\hat{Q}$ and $\hat{M}$ are defined as:
\begin{equation}
    \hat{M}=\,M/\alpha^{1/2}\,,\;\; \hat{Q}=Q/\alpha^{1/2}\,,
\end{equation}
In the bottom panels, we present the analogous results for the case $\tilde{\sigma}=0,\,\pm 0.5,\,\pm 1,\, \pm 5$ and $\tilde{\gamma}=0$. The horizontal axis in these plots corresponds to the normalized mass with respect to the GB coupling, $\hat{M}=M/\alpha^{1/2}$, with $M=0.5$.

When $r_{\rm np}$ exceeds $r_h$ part of the exterior cannot be described by the perturbative solution.  One can read the corresponding mass from Fig.~\ref{fig:perturbative_treatment}. From both top panels, we deduce that this mass increases/decreases for positive/negative values of $\tilde{\gamma}$. We also see that for $\hat{M}\gtrapprox 2.5 $ all curves start merging, as the $\gamma$-term becomes significantly subdominant with respect to the GB one. One other interesting property we notice occurs for the value of $\tilde{\gamma}=0.5$  and it pertains to more than one solution existing for the same mass which can be seen in the right panel of Fig.~\ref{fig:perturbative_treatment}. The radius of the singularity does not display similar behaviour and different mass black holes have different singularity radii, which is shown in the left panel of Fig.~\ref{fig:perturbative_treatment}. These solutions are different from one another however as they describe black holes with different scalar charges.

From the bottom panels, we notice a similar trend regarding the effects of $\tilde{\sigma}$ and the mass for which perturbativity is lost already at the exterior. For $\hat{M}\gtrapprox 3.5$ all solutions in the $\hat{M}$-$\hat{Q}$ plots begin to merge as the $\sigma$-term becomes subdominant. One of the main differences with respect to the top-panel plots, however, has to do with the maximum scalar charge. In the $\tilde{\sigma}=0,\tilde{\gamma}\ne 0$ scenario we manage to get solutions with substantially larger scalar charges in comparison to the $\tilde{\sigma}=\tilde{\gamma}=0$ (blue line). In the $\tilde{\gamma}=0,\tilde{\sigma}\ne 0$ case this does not happen. Let us also point out that the values for $r_{\text{np}}$ depicted in the bottom left panel, are the same for positive and negative values of $\sigma$. To understand why this occurs it is helpful to see the way $\sigma$ enters the perturbative expansions, which are presented in \ref{sec:appendix_perturbative}. Specifically, $\sigma$ appears at second order in the expansion of the scalar field as a multiplicative constant, which explains why changing its sign yields the same solution for $r_{\text{np}}$.

It should be noted, however, that these conclusions have to be drawn with care, as they correspond to a region of the parameter space where $\hat{M}<5$, or $\tilde{\alpha}>0.01$. This is a region that can in principle render the perturbative approach problematic in general and only a proper numerical analysis can either confirm or disprove the aforementioned effects.

\section{Numerical results}
\label{sec:numerical}
We now move to solve the full system of equations numerically. This is a system of ordinary partial differential equations (ODEs)  of the form $\{\phi'',A',B'\}=f(r,\phi',\phi,A,B)$. We separate the analysis into two regions: the black hole exterior and the black hole interior. In both cases, the  integration starts at the horizon. The theoretical parameter space consists of $(\gamma,\,\sigma,\,\kappa,\,r_h)$, where $r_h$ is the black hole horizon radius. Since $r_h$ appears in the existence condition \eqref{eq:phidh}, the allowed values for the coupling parameters are expected to be affected if we variate $r_h$. We can straightforwardly reduce the dimension of the parameter space by one if we normalize the coupling parameters with the horizon radius as we did in the previous section.

For a given theory defined by $(\tilde{\gamma},\,\tilde{\sigma},\,\tilde{\kappa})$ we allow the values of $\tilde{\alpha}$ to scan the parameter space starting from small $\tilde{\alpha}$ and gradually increasing until the existence conditions are saturated, We, therefore, need the set of values $\{\phi',\phi,A,B\}_{r_h}$. Despite appearing to constitute ``initial data'', this set of values is not entirely free to choose. In practice, in order to apply the existence conditions \eqref{eq:condition_1}-\eqref{eq:condition_2} with reasonable numerical accuracy, 
we use a perturbative expansion near the horizon and we numerically solve the system of algebraic equations for the first few coefficients appearing in the expansions (up to order $\mathcal{O}(r-r_h)^2$).  This process reduces the number of the free initial conditions to two, namely the value of the scalar field at the horizon and that of the first-order coefficient of $A$. The latter one, however, is fixed by asymptotic flatness, leaving $\phi_h$ as the only free-to-chose initial condition. The asymptotic value of the scalar field should be constant but otherwise unconstrained since our model is shift-symmetric. For simplicity, we choose $\phi_h$ so that $\phi_\infty=0$. To achieve that we employ a shooting method while integrating outwards, demanding $\phi$ vanishing to a part in $10^4$. The remaining free parameter is the horizon radius $r_h$.
We then start the numerical integration outwards (inwards) from $r=r_h\pm\mathcal{O}(10^{-5})$. In the exterior, we typically integrate up to $r/r_h\approx 10^{5}$.

In the following subsections, we present plots corresponding to different cases of couplings. In each case we numerically calculate the scalar charge and the the Arnowitt-Deser-Misner (ADM) mass of the black hole using the following expressions:
\begin{equation}
    Q=-\lim_{r\rightarrow\infty}\left(r^2\phi'\right)\,,\quad M=\lim_{r\rightarrow \infty}\left[\frac{r \left(2-2 B+r^2 \phi'^2\right)}{r^2 \phi'^2-4}\right].
\end{equation}
We were also able to verify the emergence of a finite-radius singularity, consistent with the existence conditions. While integrating from the horizon and inwards we noticed the following general trend: Starting from GR ($\alpha\rightarrow 0$) and gradually increasing the couplings, the geometric invariants diverge and the solutions become singular at some radius $r_s$. The larger GR deviations become the more $r_s$ approaches $r_h$. When one of the existence conditions is saturated the singularity radius approaches the horizon radius, i.e. $r_s\rightarrow r_h$.

\subsection{Charge, mass and scalar profile}
\begin{figure}[ht]
    \centering
    \includegraphics[width=.47\linewidth]{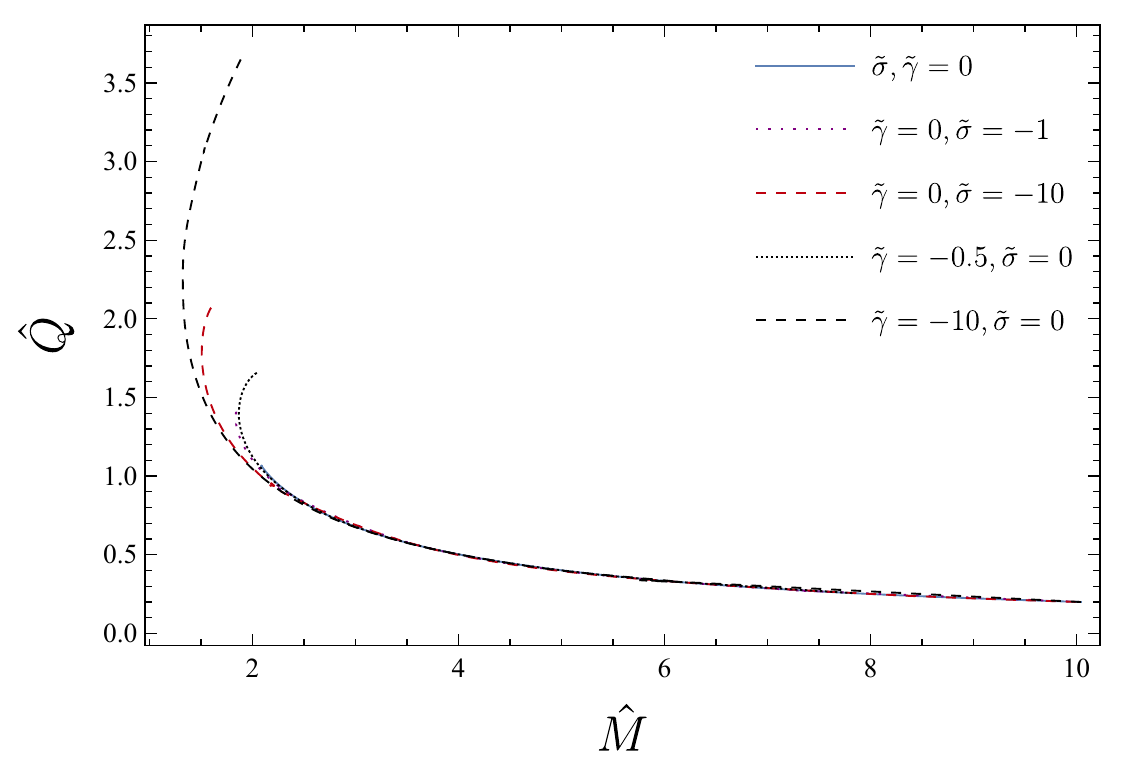}
    \hspace{5mm}
    \includegraphics[width=.47\linewidth]{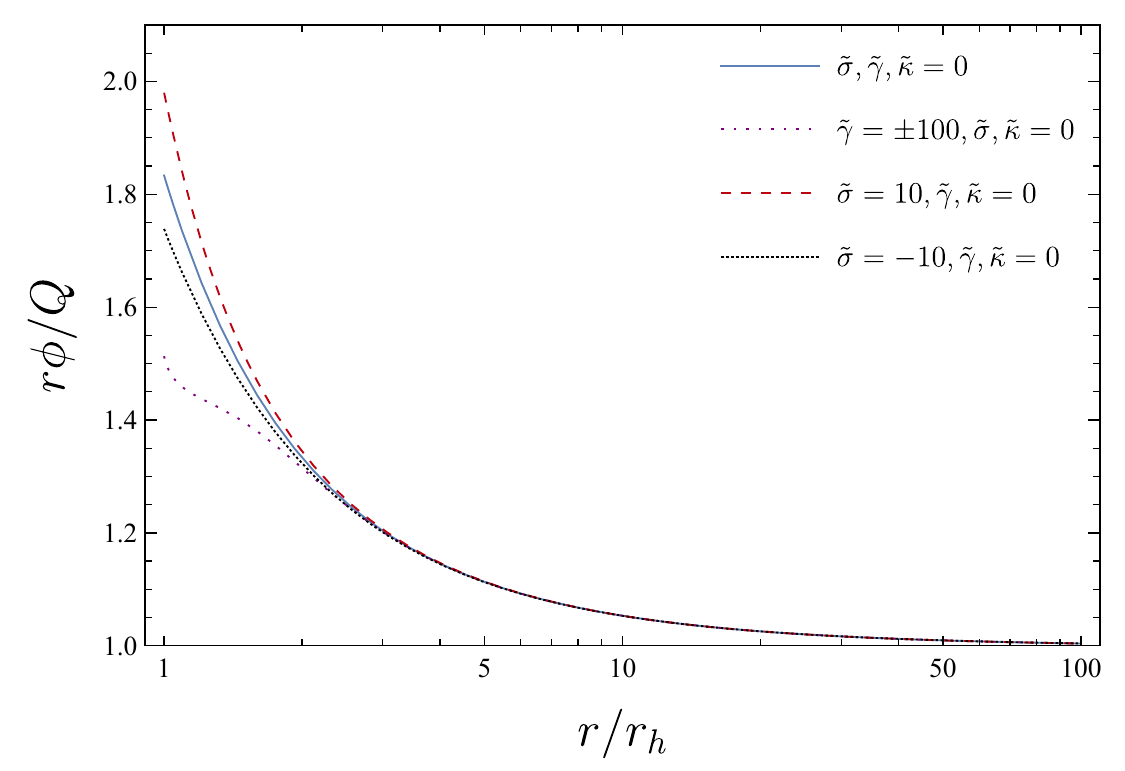}
    \caption{\textit{Left:} The relation between the normalized mass and charge. \textit{Right:} The scalar field profile for black hole solutions with $\hat{M} \sim 10$.}
    \label{fig:MQphi}
\end{figure}
In these subsections we attempt to present the overall generic trend that the black hole properties follow, if one considers action \eqref{eq:action}. We discuss the charge, mass and scalar profile for a few examples corresponding to different scenarios, which motivate the more thorough analysis that follows in the next subsections.

On the left panel of Fig.~\ref{fig:MQphi}, we show the $\hat{M}$-$\hat{Q}$ plot for different negative values of the coupling constants $\tilde{\gamma}$ and $\tilde{\sigma}$. The corresponding plots for positive couplings are not presented here, since -at these scales- they are overshadowed by the $\tilde{\gamma} = \tilde{\sigma} = 0$ curve, as will be explained later in more detail. In all positive-coupling cases, the minimum black hole mass is larger than the one corresponding to $\tilde{\gamma} = \tilde{\sigma} = 0$. Furthermore, non-zero $\tilde{\kappa}$ curves are almost indistinguishable from the $\tilde{\gamma} = \tilde{\sigma} = 0$ one since as we saw $\tilde{\kappa}$ does not enter the existence conditions. Consequently, the corresponding $\tilde{\kappa}$-plots are not presented here.
We see that for large $\hat{M}$ which corresponds to small GB couplings, the charge in all cases drops off to zero and GR is retrieved. This is of course associated with the fact that the GB term is the one sourcing the hair. In the small $\hat{M}$ regime significant deviations are observed, which are explained in the following coupling-specific subsections. On the right panel of Fig.~\ref{fig:MQphi}, we show the profile of the scalar field, properly normalized with the distance and the scalar charge. All curves exhibit a $1/r$ fall-off and asymptotically approach $1$. For large radii, the scalar field profiles are indiscernible for different couplings. In the near horizon regime, however, there are apparent deviations in accordance with the non-trivial deviations shown in the left panel.

To make things easier for the reader, in what follows, we consider the GB coupling $\tilde{\alpha}$ in combination with $\tilde{\gamma},\,\tilde{\sigma}$ and $\tilde{\kappa}$ separately. In this work, we consider $\alpha>0$ as this is consistent with most of the bibliography. However, it is worth pointing out, that action \eqref{eq:action} is invariant under the simultaneous transformation $\alpha\rightarrow-\alpha$, $\phi\rightarrow-\phi$ and $\sigma\rightarrow-\sigma$ and that in the case of $\sigma=\gamma=\kappa=0$, the sign of $\phi$ is determined by the sign of $\alpha$ for solutions that are continuously connected to Schwarzschild as $\alpha \to 0$. In what follows we consider both positive and negative values for $\sigma$, $\gamma$, and $\kappa$, and hence our analysis should effectively cover the $\alpha<0$ case as well, at least for configurations that are continuously connected to Schwarzschild.

\subsection{The $\tilde{\gamma}$ term}

\begin{figure}[t]
    \centering
    \includegraphics[width=.48\linewidth]{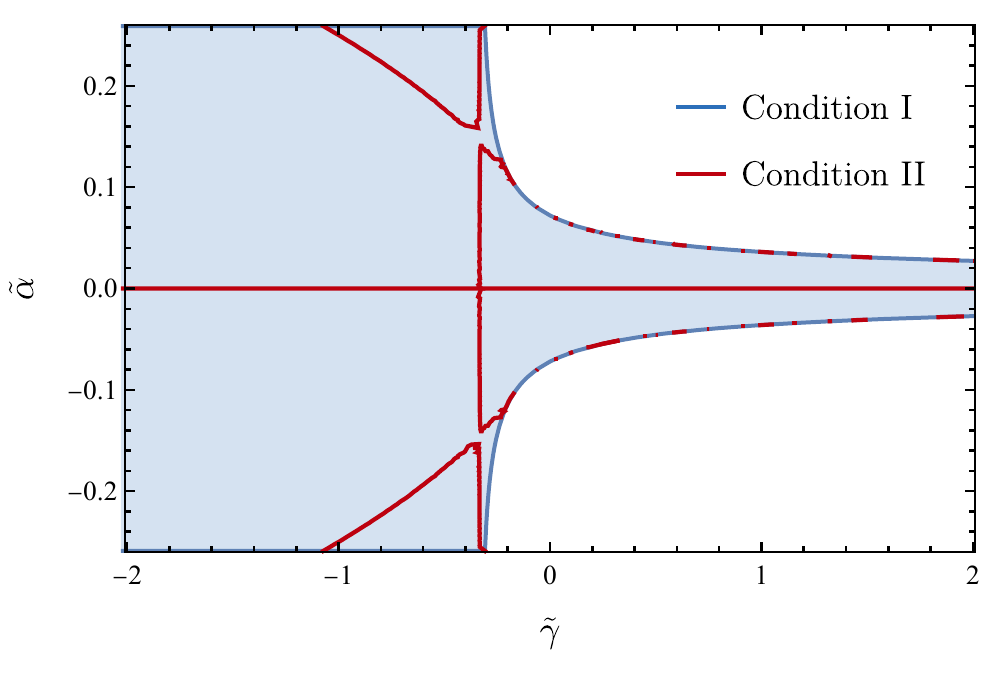}\hspace{5mm}
    \includegraphics[width=.47\linewidth]{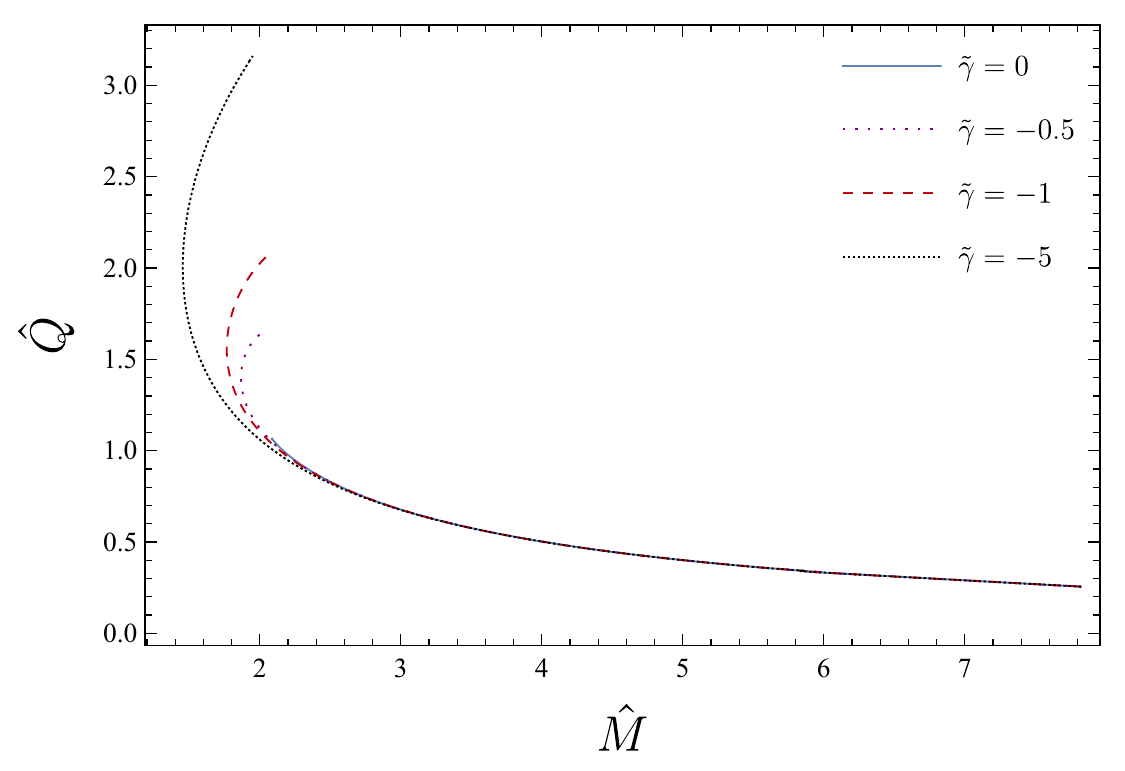}
    
    \caption{\textit{Left}: Existence conditions in the case of $\tilde{\sigma}=\tilde{\kappa}=0$. The blue shaded region corresponds to the area of the parameter space allowed by condition I \eqref{eq:condition_1}. The red line corresponds to the values within the allowed blue region that are excluded by condition II \eqref{eq:condition_2}. \textit{Right}: Mass-Charge plots for $\tilde{\sigma} = \tilde{\kappa} = 0$ and $\tilde{\gamma}=\{-5,-1,-0.5,0\}$.}
    \label{fig:existence_QM_gamma}
\end{figure}

First we consider the case $\tilde{\sigma}=\tilde{\kappa}=0$. From \eqref{eq:condition_1} and \eqref{eq:condition_2} we find the conditions on $\tilde{\gamma}$ necessary for regularity at the horizon. The existence conditions I-II are in general non-trivial and the easiest way to track them is to examine the corresponding region plot. In the left panel of Fig.~\ref{fig:existence_QM_gamma} we see the aforementioned plot with $\tilde{\gamma}$ being on the horizontal axis and $\tilde{\alpha}$ occupying the vertical one. The first obvious observation relates to the apparent asymmetry about the vertical axis. Therefore, we expect the sign of $\tilde{\gamma}$ to influence the black hole solutions and properties. In particular, for negative values of $\tilde{\gamma}$ the parameter space of allowed values for $\tilde{\alpha}$ increases, and so we expect negative values of $\tilde{\gamma}$ to allow for hairy solutions with smaller masses. On the other hand when $\tilde{\gamma}>0$ the parameter space of $\tilde{\alpha}$ shrinks and we expect the black hole mass range to also decrease. Regarding the GB-coupling $\tilde{\alpha}$, extending the plot to negative values of $\tilde{\alpha}$ is trivial as, for $\tilde{\sigma}=0$, the action \eqref{eq:action} is invariant under the simultaneous transformation $\tilde{\alpha}\rightarrow-\tilde{\alpha}$ and $\phi\rightarrow-\phi$.

\begin{figure}[t]
    \centering
    \includegraphics[width=.47\linewidth]{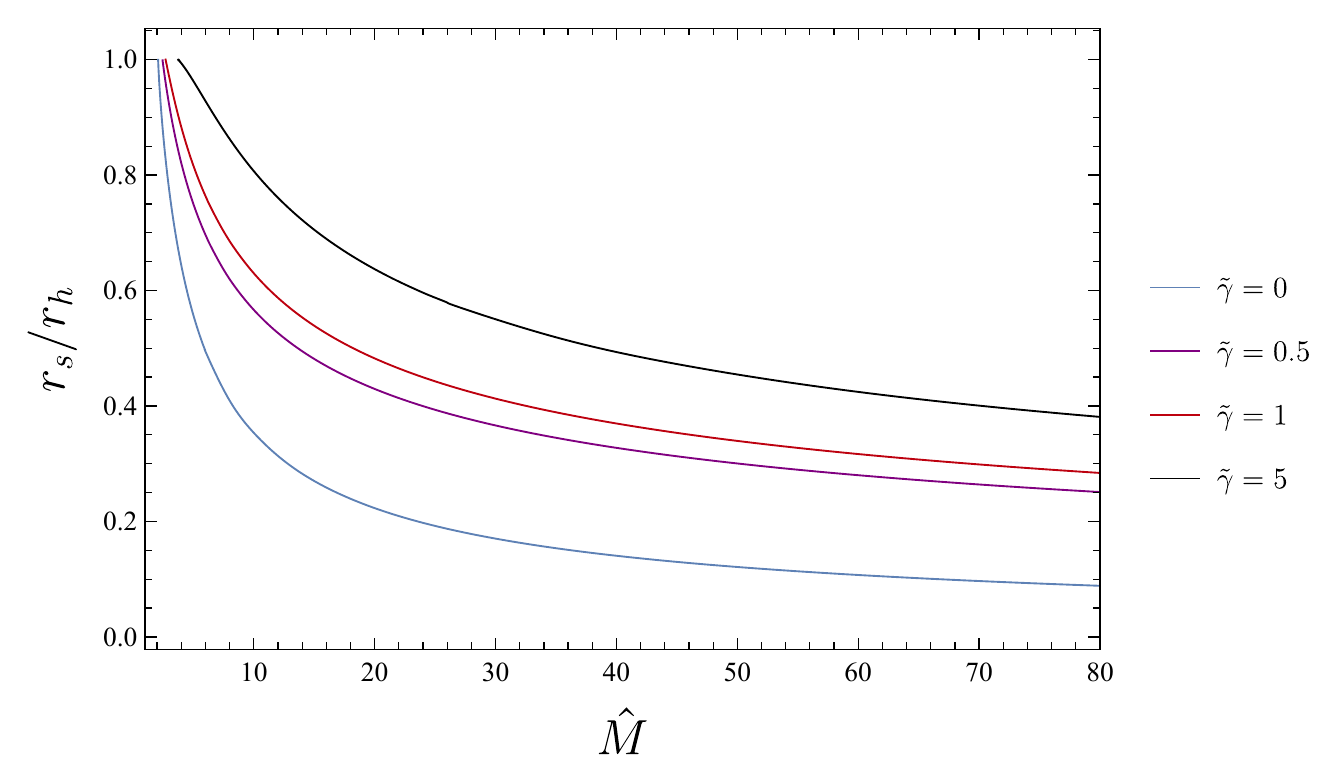}\hspace{5mm}
    \includegraphics[width=.47\linewidth]{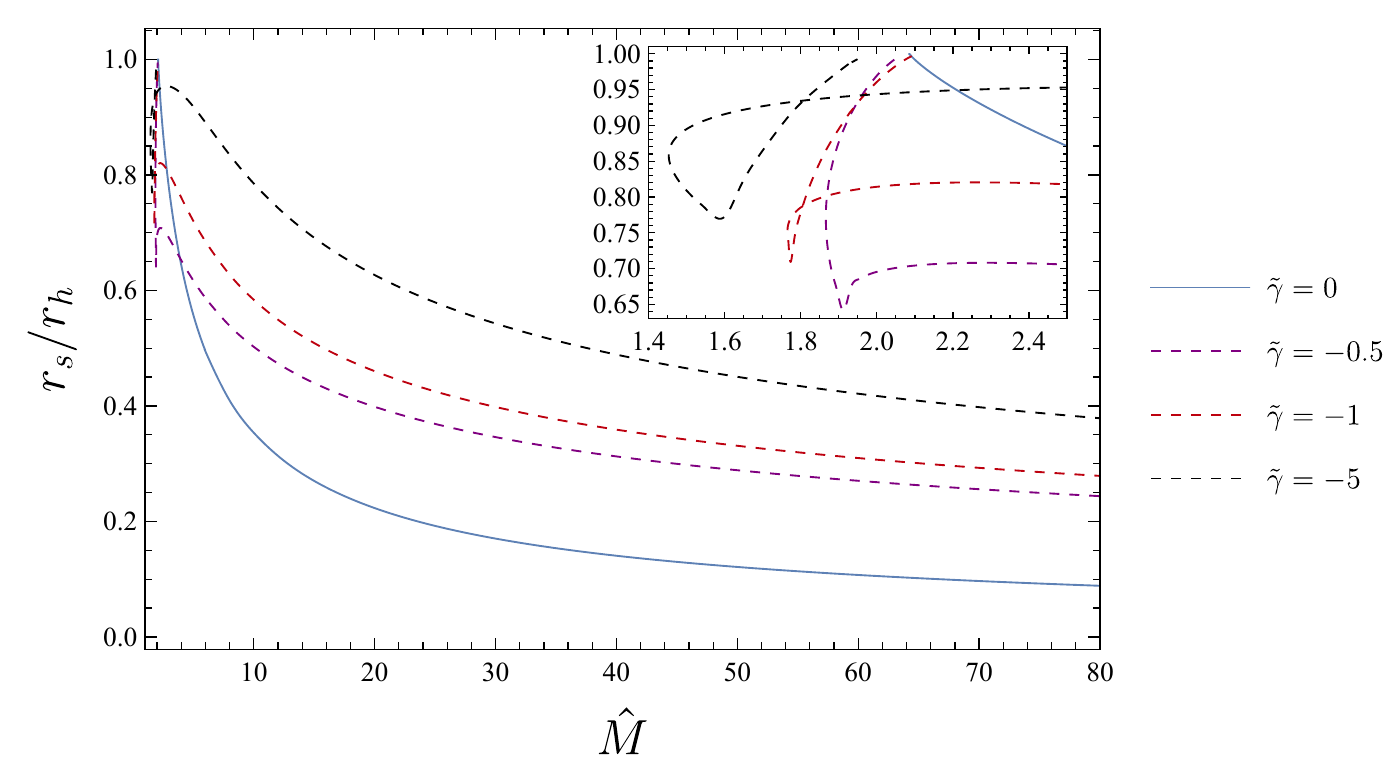}
    \caption{The finite singularity radius $r_{s}$ as a function of the normalized mass $\hat{M}$ for $\tilde{\sigma} = \tilde{\kappa} = 0$ and $\tilde{\gamma}=\{-5,-1,-0.5,0,0.5,1,5\}$.}
    \label{fig:singularity_gamma}
\end{figure}

These are indeed verified in Fig.~\ref{fig:singularity_gamma}, where the emergence of a finite-radius singularity is demonstrated in the interior of the black hole. The left panel shows the singularity radius of the black hole mass for $\tilde{\gamma}=\{0,0.5,1,5\}$ while the right panel shows the corresponding results for $\tilde{\gamma}=\{0,-0.5,-1,-5\}$. The values are chosen to be of order $\sim 1-10$ with respect to $\tilde{\alpha}_{\text{max}}$, where $\tilde{\alpha}_{\text{max}}$ corresponds to the largest allowed value for $\alpha$ satisfying the existence conditions.  For the choices of $\tilde{\gamma}$ made, we present the results for the minimum hairy black hole mass in the following table:

\begin{table}[h]
    \centering
    \begin{tabular}{c||c|c|c|c|c|c|c}
    \multicolumn{8}{c}{Minimum mass for $\tilde{\sigma}=\tilde{\kappa}=0,\; \tilde{\gamma} \ne 0,\,\tilde{\alpha}>0$}\\ \hline
    \raisebox{-1mm}{$\tilde{\gamma}$} & \raisebox{-1mm}{$-5.0$} & \raisebox{-1mm}{$-1.0$} & \raisebox{-1mm}{$-0.5$} & \raisebox{-1mm}{\;\,$0.0$\;\,} & \raisebox{-1mm}{$+0.5$} & \raisebox{-1mm}{$+1.0$} & \raisebox{-1mm}{$+5.0$} \\[2mm] \hline
    \raisebox{-1mm}{$\hat{M}$} & 1.45 & 1.77 & 1.87 & 2.08 & 2.45 & 2.71 & 3.75 \\[2mm]
    \end{tabular}
\end{table}

For a negative $\tilde{\gamma}$, we notice another interesting property of the solutions: at small masses, the apparent change in monotonicity in the $\hat{M}$-$\hat{Q}$ and $\hat{M}$-$r_s$ (see the inset) plots indicates that black holes with the same mass can correspond to different scalar charges and singularity radii. Therefore, one would expect that the black hole with the larger scalar charge, would shed some of it to reach a more favourable scalar configuration with a smaller charge. Finally, from Fig.~\ref{fig:singularity_gamma} it is pointed out that in the larger mass regime, the sign of $\tilde{\gamma}$ becomes unimportant and the cases with opposite signs merge.

\subsection{The $\tilde{\sigma}$ term}

\begin{figure}[t]
    \centering
    \includegraphics[width=.47\linewidth]{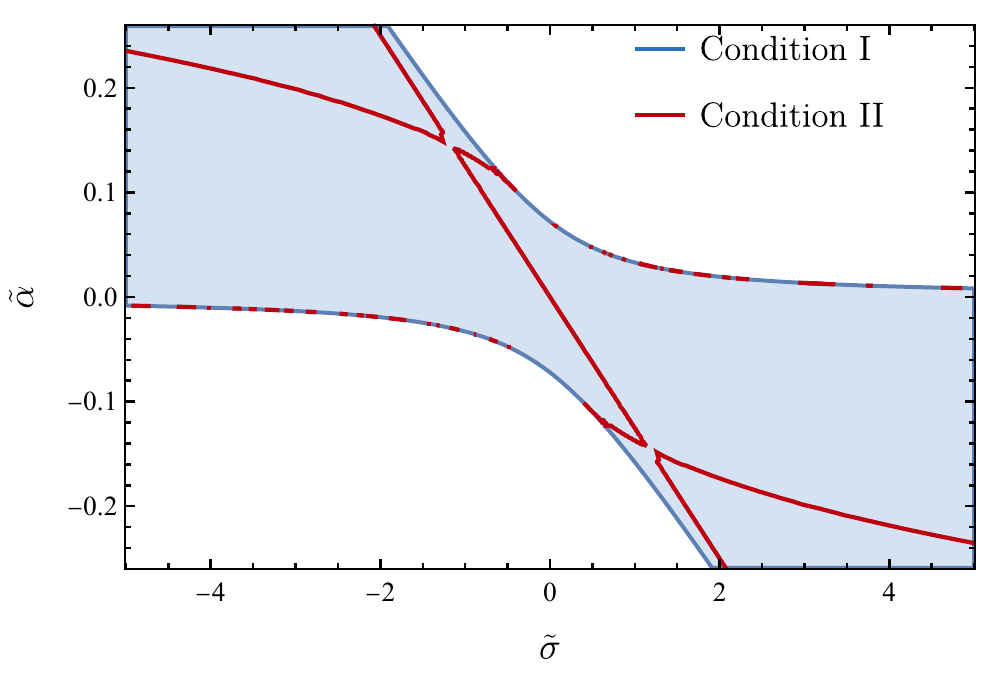}\hspace{5mm}
    \includegraphics[width=.47\linewidth]{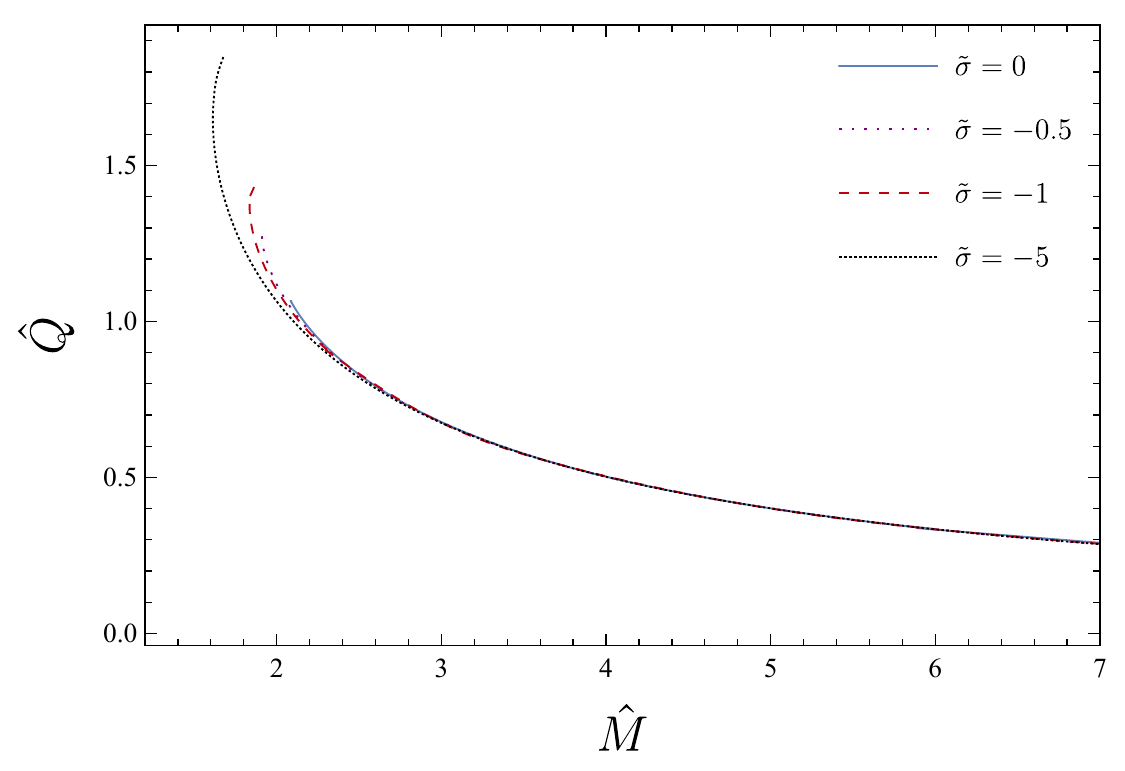}
    
    \caption{\textit{Left}: Existence conditions in the case of $\tilde{\gamma}=\tilde{\kappa}=0$. The blue shaded region corresponds to the inequality of condition I \eqref{eq:condition_1}, while the red line corresponds to the inequality of condition II \eqref{eq:condition_2}. \textit{Right}: Mass-Charge plots for $\tilde{\gamma} = \tilde{\kappa} = 0$ and $\tilde{\sigma}=\{-5,-1,-0.5,0\}$.}
    \label{fig:existence_QM_sigma}
\end{figure}

In the left panel of Fig. \ref{fig:existence_QM_sigma} we present the allowed and excluded regions of the parameter space according to the existence conditions in the case of $\tilde{\gamma} = 0$, with $\tilde{\sigma}$ being on the horizontal and $\tilde{\alpha}$ on the vertical axis. For negative values of $\tilde{\alpha}$ the region plot we retrieve demonstrates an origin symmetry which was anticipated since the action \eqref{eq:action} is invariant under the simultaneous transformation $\tilde{\alpha}\rightarrow-\tilde{\alpha}$, $\tilde{\sigma}\rightarrow-\tilde{\sigma}$, and $\phi\rightarrow-\phi$.
For $\tilde{\alpha}>0,\,\tilde{\sigma}<0$ the allowed values for $\tilde{\alpha}$ increase and therefore the mass range also increases, and hairy black holes with smaller masses are found. At the same time for $\tilde{\alpha}>0,\,\tilde{\sigma}>0$, the parameter space of $\tilde{\alpha}$ shrinks and the black hole mass range should also decrease. If we considered $\tilde{\alpha}<0$ the above conclusions would be reversed.

\begin{figure}[t]
    \centering
    \includegraphics[width=.47\linewidth]{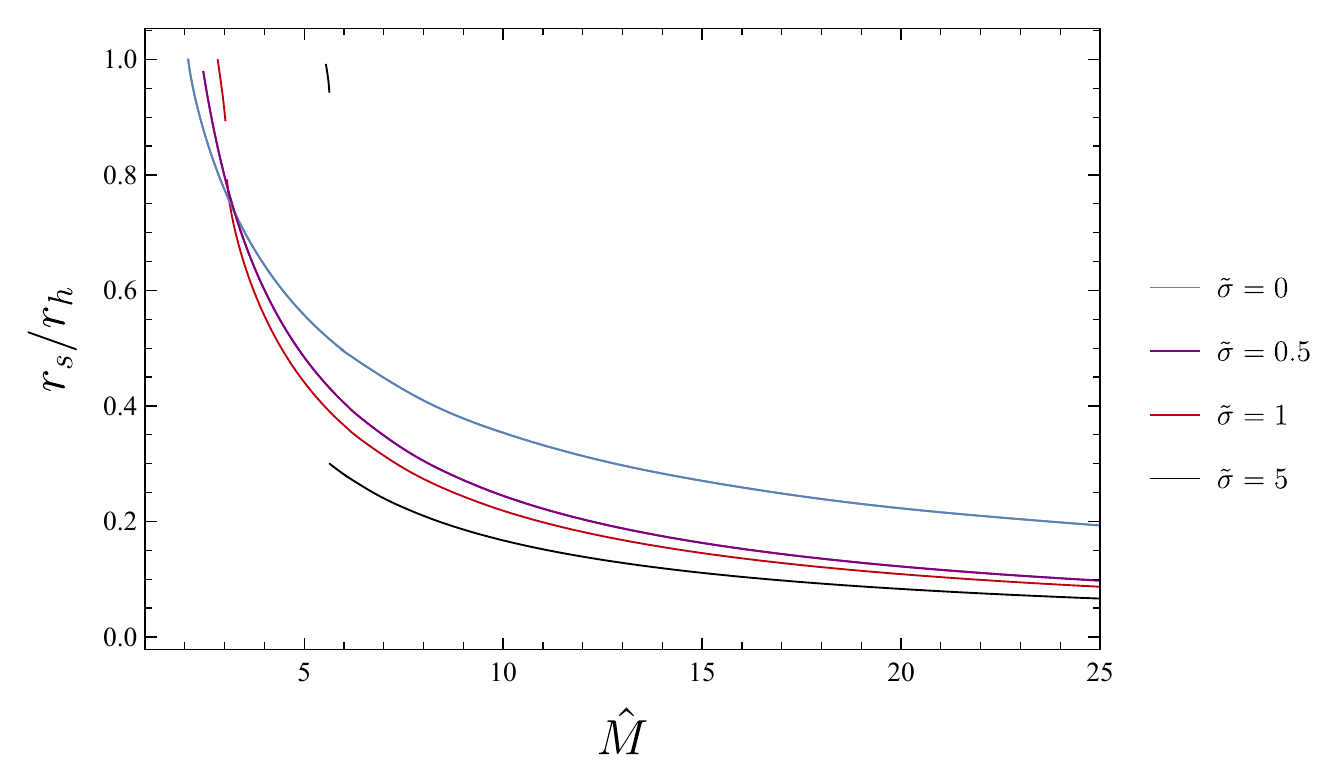}\hspace{5mm}
    \includegraphics[width=.47\linewidth]{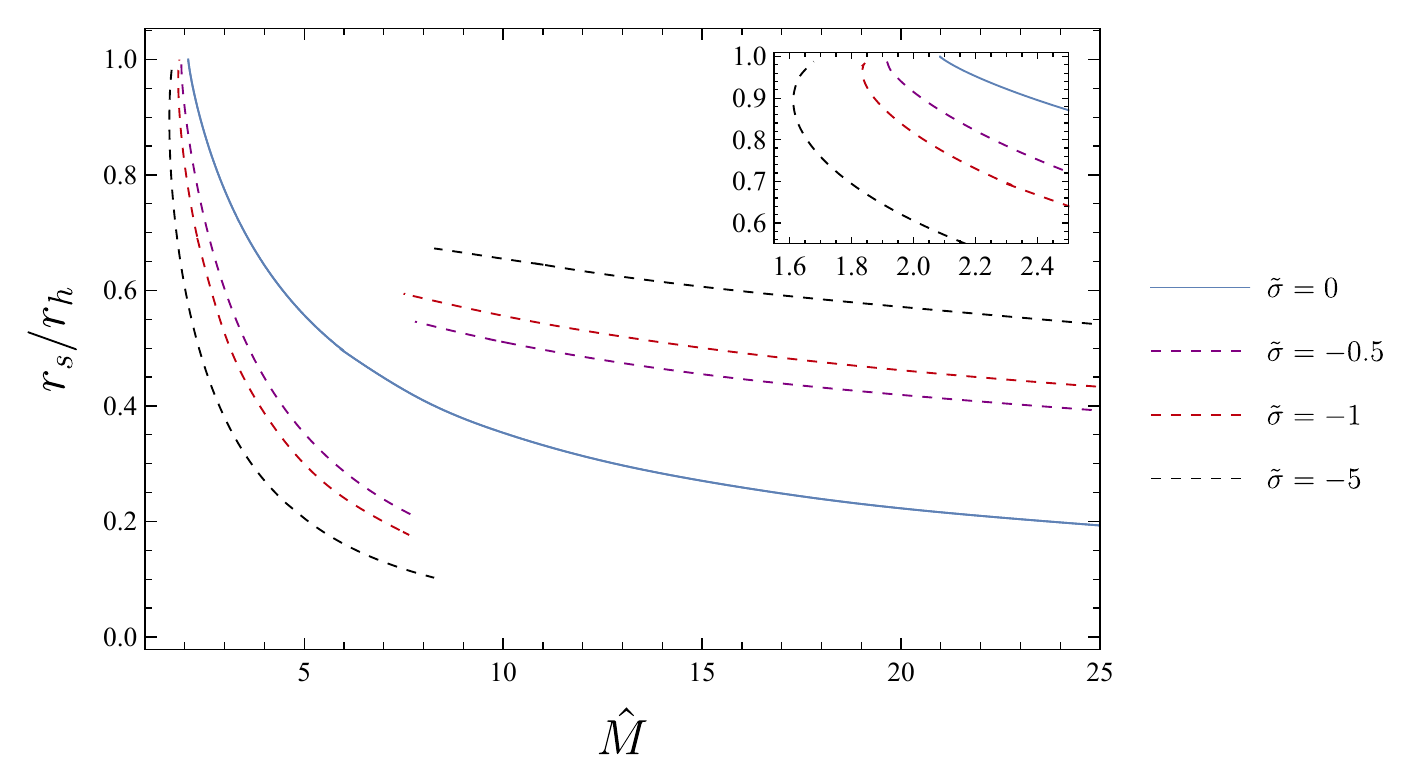}
    \caption{The finite singularity radius $r_{s}$ as a function of the normalized mass $\hat{M}$, for $\tilde{\gamma} = \tilde{\kappa} = 0$. The horizon radius $r_h = 1$.}
    \label{fig:singularity_sigma}
\end{figure}

In Fig. \ref{fig:singularity_sigma} we display the singularity radius in this scenario and its dependence on the value of $\tilde{\sigma}$. Verifying the above, positive(negative) $\tilde{\sigma}$ leads to a larger(smaller) minimum black hole mass.
\begin{table}[h]
    \centering
    \begin{tabular}{c||c|c|c|c|c|c|c}
    \multicolumn{8}{c}{Minimum mass for $\tilde{\gamma}=\tilde{\kappa}=0,\; \tilde{\sigma} \ne 0\,,\tilde{\alpha}>0$}\\ \hline
    \raisebox{-1mm}{$\tilde{\sigma}$} & \raisebox{-1mm}{$-5.0$} & \raisebox{-1mm}{$-1.0$} & \raisebox{-1mm}{$-0.5$} & \raisebox{-1mm}{\;\,$0.0$\;\,} & \raisebox{-1mm}{$+0.5$} & \raisebox{-1mm}{$+1.0$} & \raisebox{-1mm}{$+5.0$} \\[2mm] \hline
    \raisebox{-1mm}{$\hat{M}$} & 1.61 & 1.84 & 1.91 & 2.08 & 2.47 & 2.83 & 5.55  \\[2mm]
    \end{tabular}
\end{table}

In the $\tilde{\sigma}\ne 0$ scenario, the relation between the finite singularity radius and the normalized mass exhibits discontinuous behaviour, which is evident from the vertical jumps shown in Fig. \ref{fig:singularity_sigma}. As already explained, we identify the singularity radius as the one for which a geometric invariant (e.g., the Gauss-Bonnet or equivalently the Kretschmann invariant) diverges. To explain the discontinuity let us imagine that we start from some large $\hat{M}$ moving inwards towards smaller masses. At $r=r_s$ the GB invariant diverges and we identify $r_s$ as the singularity radius. There exists a second special point at $r=r_s'>r_s$ where the metric functions and the scalar field appear to lose differentiability because they develop a kink. This appears to be because the second derivative become discontinuous.
The differential equations however can still be integrated for $r_s'>r>r_s$. If we plotted $r_s'$ instead of $r_s$, then the vertical jump would no longer be present and the lines would be continuous. In all cases, however, we chose to plot the singularity corresponding to the divergence of the geometric invariants. On the other hand, for positive $\tilde{\sigma}$, we do not encounter any other ``singularities'' than the ones we plot, which correspond once again to the geometric invariants diverging.

Similar discontinuities have also been encountered in scalar Einstein-scalar-Gauss-Bonnet gravity with a quadratic exponential coupling \cite{2022arXiv220710692F}. In the zoomed-in part of the right panel of Fig. \ref{fig:singularity_sigma}, similar behaviour to the negative $\tilde{\gamma}$ case is exhibited, where same-mass black holes have different singularity radii. This can also be understood from the $\hat{M}$-$\hat{Q}$ plot, in the right panel of Fig.~\ref{fig:existence_QM_sigma}, where a turning point appears at small masses.

\subsection{The $\tilde{\kappa}$ term}
\begin{figure}[t]
     \centering
    \includegraphics[width=.47\linewidth]{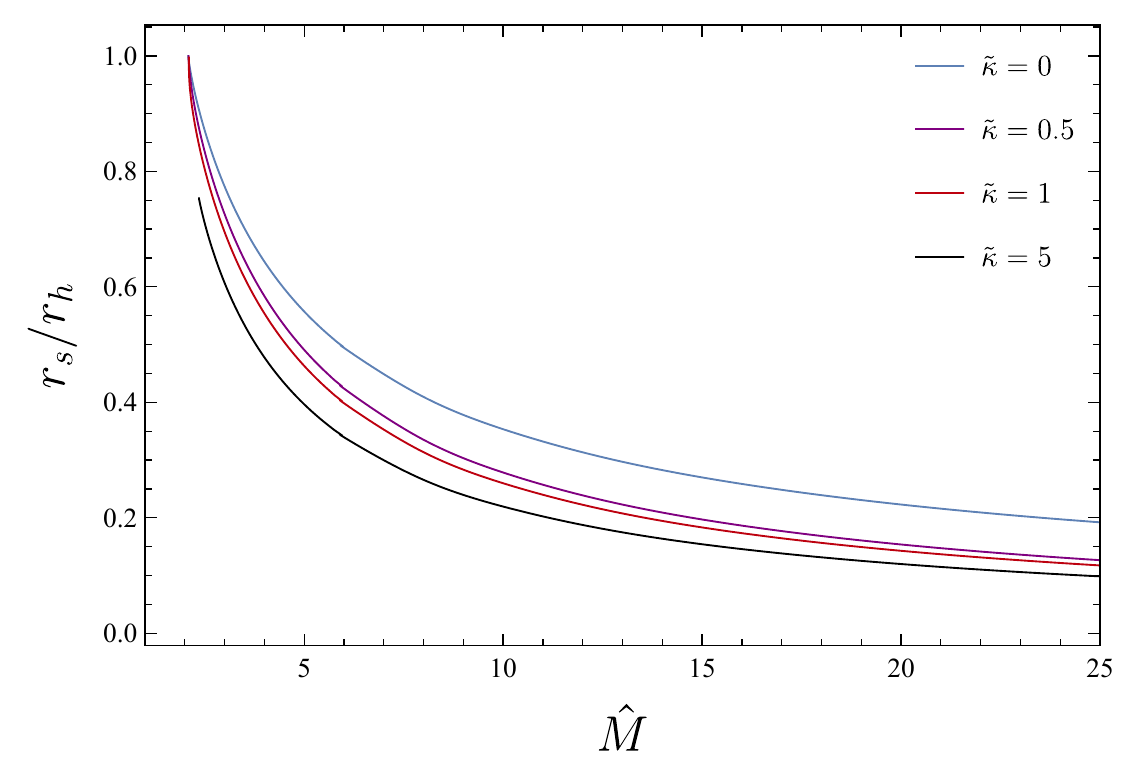}
    \hspace{5mm}
    \includegraphics[width=.47\linewidth]{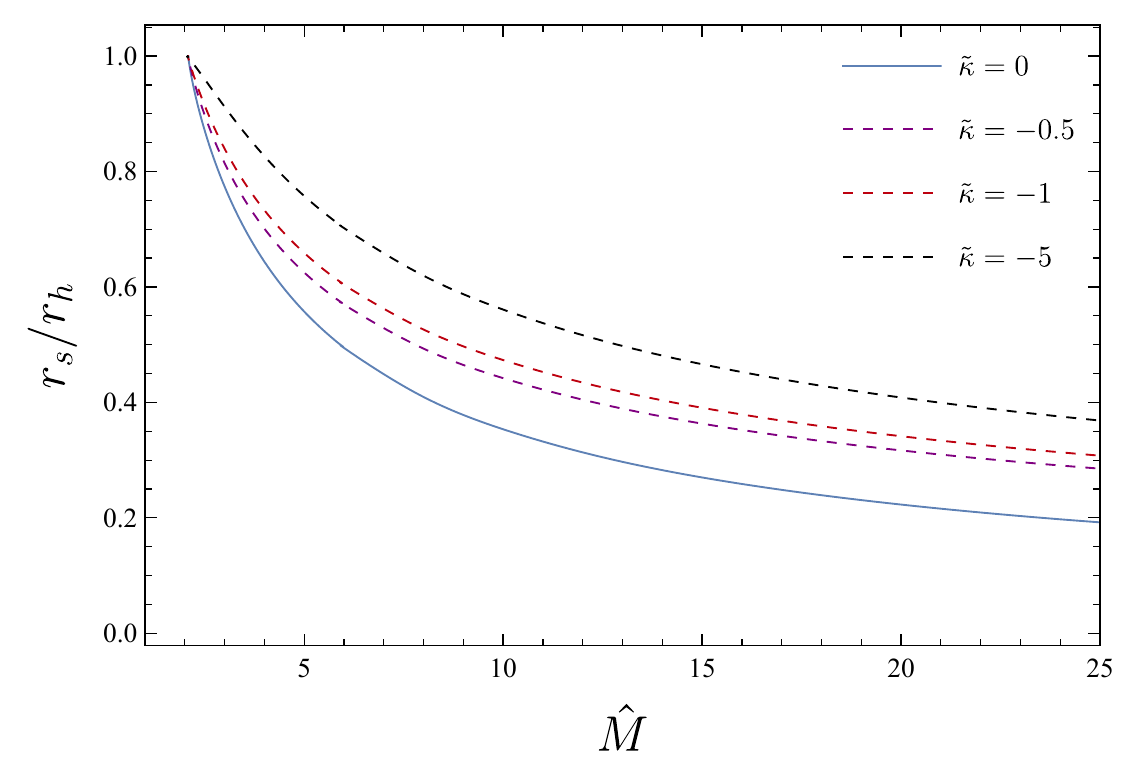}
    \caption{The finite singularity radius $r_{s}$ as a function of the normalized mass $\hat{M}$, with $\tilde{\sigma} = \tilde{\gamma} = 0$. The horizon radius $r_h = 1$.}
    \label{fig:singularity_kappa_f}
\end{figure}

It is evident from equations \eqref{eq:condition_1} and \eqref{eq:condition_2} that  $\kappa$ does not enter the existence conditions. As a result, one might naively conclude that black hole solutions exist irrespective of the value that $\kappa$ takes, given that the remaining parameters satisfy the existence conditions. Contrarily, that is not the observed behaviour. If $\kappa$ is taken to be positive, then we cannot  find solutions for all values of $\alpha$ that are allowed by the existence conditions; however, if $\kappa$ is negative, then solutions could be found for all values of $\alpha$ allowed by the conditions. This behaviour is illustrated in Fig. \ref{fig:singularity_kappa_f}, where for negative values of $\kappa$ it is possible to saturate the existence condition and have solutions with a naked singularity, but for $\kappa > 0$, in general, that cannot be achieved.  
To better understand this trend, it is useful to rewrite the scalar equation for $\gamma = \sigma = 0$ as
\begin{align}
h^{\mu \nu} \nabla_{\mu}\nabla_{\nu}\phi \equiv \left[g^{\mu \nu}\left(1-\kappa(\nabla\phi)^{2}\right) -2\kappa \nabla^{\mu}\phi \nabla^{\nu}\phi\right] \nabla_{\mu}\nabla_{\nu}\phi
 = -\alpha \mathcal{G}.
\label{eq:scalar_kappa}
\end{align}

\begin{figure}[t]
     \centering
    \includegraphics[width=.47\linewidth]{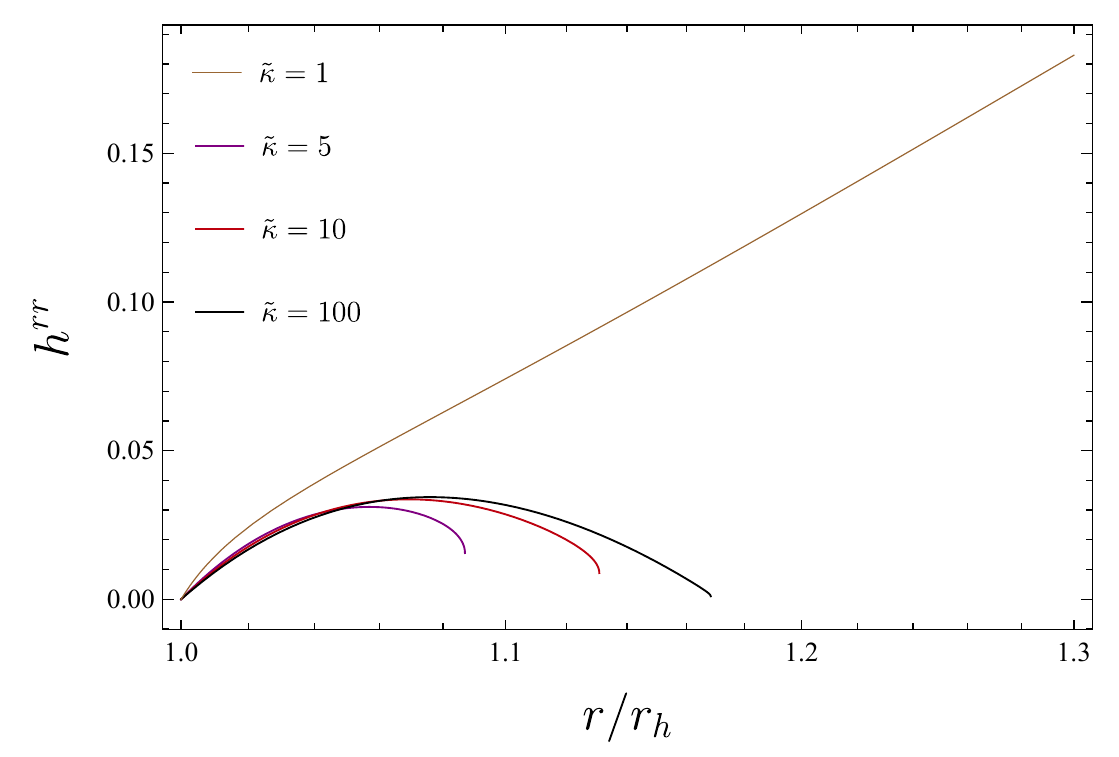}
    \hspace{5mm}
    \includegraphics[width=.47\linewidth]{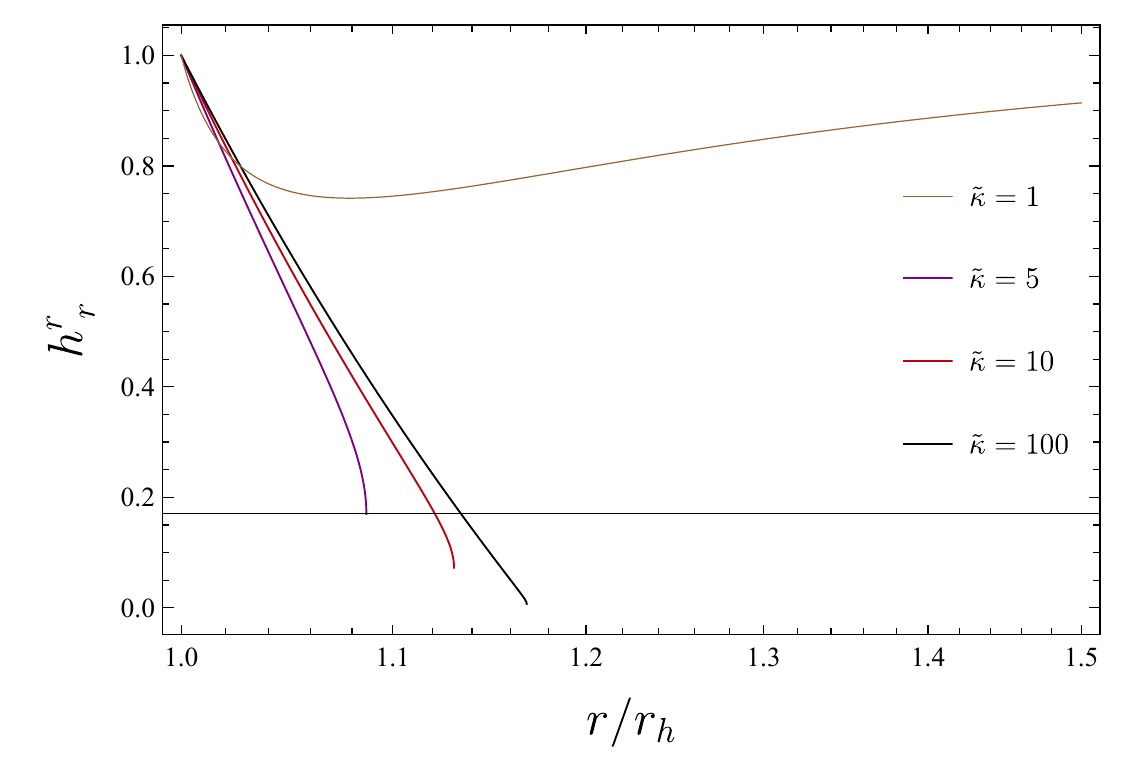}
    \caption{This figure shows the behaviour of $h_{\mu \nu}$, as defined in equation \eqref{eq:scalar_kappa}, near the horizon. \textit{Left}: $h^{rr} = B(1-3\kappa B \phi'^{2})$. \textit{Right}: $h^{r}_{\,\,r} = 1-3\kappa B \phi'^{2}$.}
    \label{fig:effective_metric}
\end{figure}

In practice, we see that when $\tilde{\kappa}>1$ not all values of $\tilde{\alpha}$, allowed by the existence conditions, yield black hole solutions. In order to give an explanation to this issue, we numerically examine the value of the quantity inside the square brackets in eq.~\eqref{eq:scalar_kappa}, namely $h^{\mu\nu}$. Due to the symmetry of our problem only $h^{rr}$ will be examined.
In Fig. \ref{fig:effective_metric}, we plot $h^{rr}$ (and $h^r_{\; r}$) for values of $\tilde{\kappa}$ spanning a few orders of magnitude, i.e. $\mathcal{O}(1)-\mathcal{O}(10^2)$. We see that for $\mathcal{O}(\tilde{\kappa})>\mathcal{O}(1)$, the quantity $h^{rr}$ approaches zero at some intermediate radius, which seems to increase as we increase the value of $\kappa$. Beyond that point, the ODE system can no longer be integrated.
This bares similarities with the behaviour of $\phi''$ at the horizon, the regulation of which yielded the existence conditions. Thus, it appears that imposing regularity for the scalar field at the horizon results in divergences appearing elsewhere for large positive values of $\tilde{\kappa}$.

\subsection{Numerical solutions vs perturbative solutions}

\begin{figure}[ht]
    \centering
    \includegraphics[width=.47\linewidth]{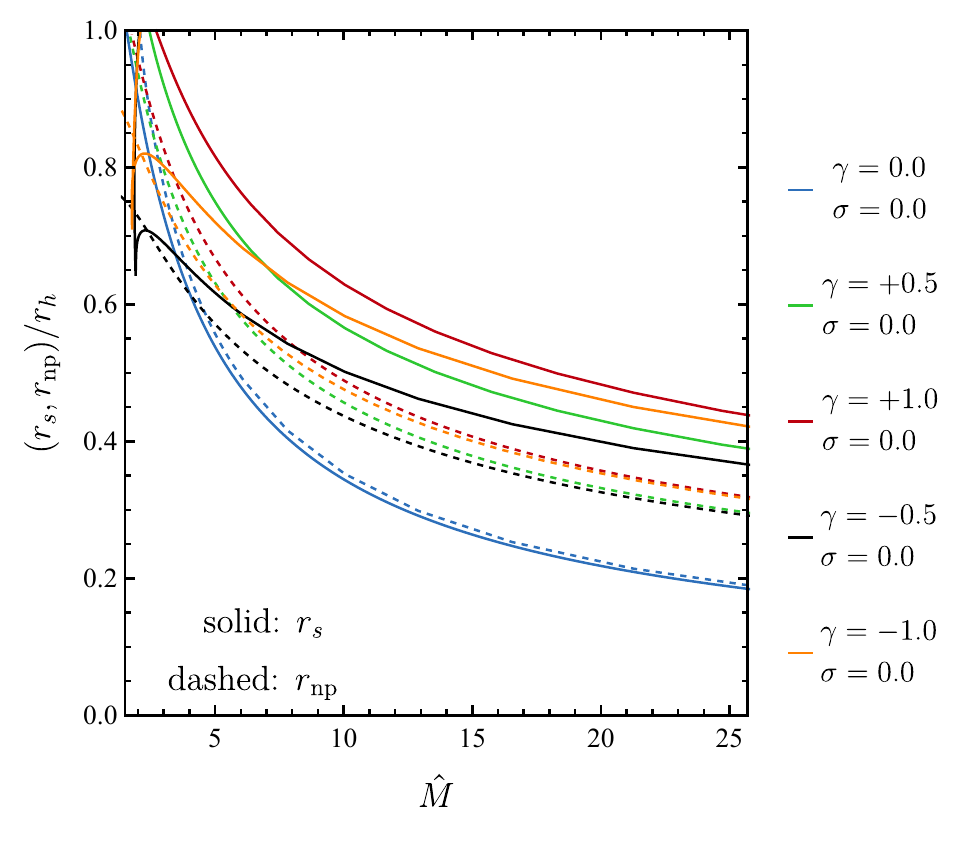}\hspace{2mm}
    \includegraphics[width=.47\linewidth]{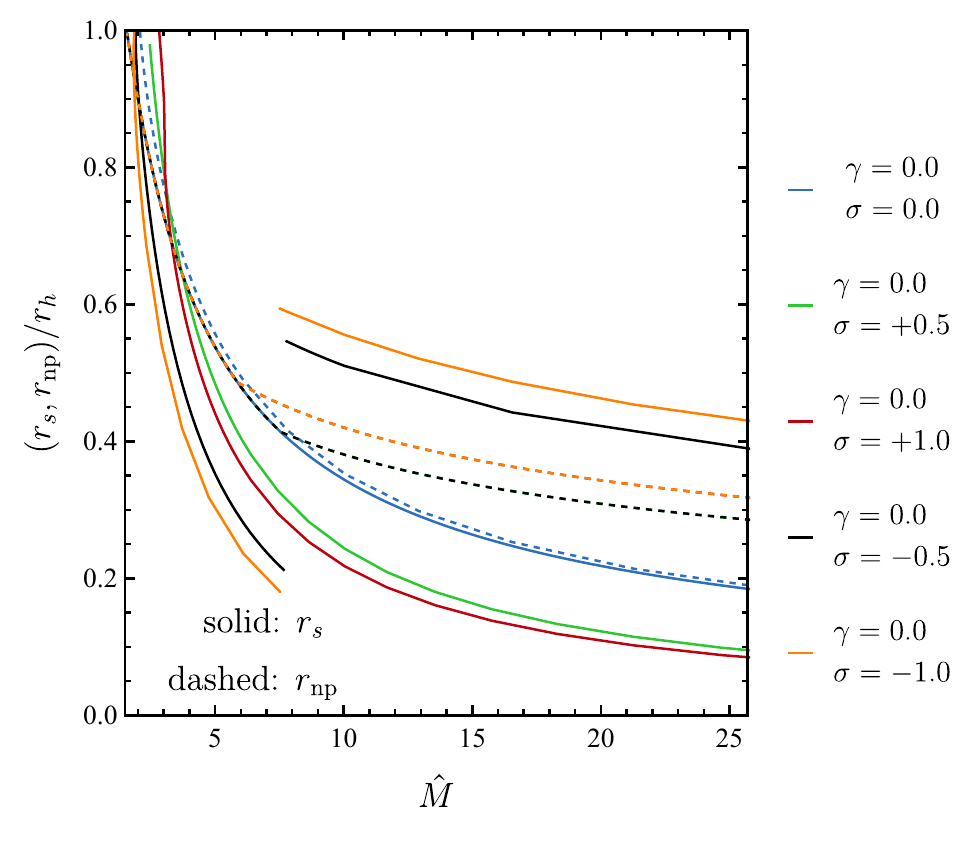}
    \label{fig:singularity_comparisons}
    \caption{\textit{Left}: Plot for the radii $r_{\text{np}}$ and $r_s$ for different values of $\gamma$ with $\sigma=0$. \textit{Right}: Same as left but for different values of $\sigma$ with $\gamma=0$.}
    \label{fig:comparisons}
\end{figure}

As mentioned earlier, it has already been demonstrated that in the case $\gamma=\sigma=0$ loss of perturbativity is associated with the appearance of a finite-radius singularity in the black hole interior. 
Here we discuss the relation between the perturbative treatment breakdown radius $r_{\text{np}}$ and the finite-radius singularity $r_s$ in the general case where $\tilde{\gamma},\,\tilde{\sigma}$ are nonzero.
We present the comparative plots in Fig.~\ref{fig:comparisons}.
Verifying the results of \cite{Sotiriou:2014pfa}, we see that the radius of the singularity in the black hole interior in the case $\tilde{\gamma}=\tilde{\sigma}=0$, is traced almost perfectly by the perturbative analysis. However, this is not the case, at least to the same level of success, when one considers the $\gamma$ and $\sigma$ contributions. From the left panel of Fig.~\ref{fig:comparisons}, we see that when $\tilde{\gamma}\ne 0$ the $r_{\text{np}}$ curve sits below the singularity radius $r_s$. From the right panel, we notice that in the $\tilde{\sigma}\ne 0$ case on the other hand the $r_{\text{np}}$ curve sits between the disconnected branches of the numerical solutions.

We can also compare the results regarding the scalar charge and mass of the black holes, by inspecting Figs.~\ref{fig:perturbative_treatment} and \ref{fig:existence_QM_gamma}, \ref{fig:existence_QM_sigma}. Specifically, for the $\tilde{\gamma}\ne 0$ case, by comparing Figs.~\ref{fig:perturbative_treatment} and \ref{fig:existence_QM_gamma} we deduce that the perturbative approach captures at least quantitatively the two main conclusions drawn by the numerical analysis: (i) For $\tilde{\gamma}>0$ the minimum black hole mass for each $\tilde{\gamma}$ increases and the mass-parameter space of solutions shrink. (ii) For $\tilde{\gamma}>0$ the minimum black hole mass for each $\tilde{\gamma}$ decreases, and for small masses it is possible to retrieve black holes of the same mass with different scalar charges. Similar conclusions were also drawn in the numerical analysis $\tilde{\sigma}\ne 0$ scenario (with the addition of the discontinuities), where the qualitative trends were also captured by the perturbative analysis.

\section{Conclusions}
\label{sec:conclusions}

We have studied hairy black holes in generalized scalar-tensor theories that exhibit a range of shift-symmetric derivative interactions, in addition to  the linear coupling to the Gauss-Bonnet invariant that is known to introduce black hole hair. We found that, although these additional interactions cannot introduce hair themselves, they can significantly influence the behaviour of the scalar fields near the horizon of the black hole and hence affect the configuration in general, including the value of a scalar charge for a given mass.

Interestingly $G_{\mu\nu}\nabla^\mu\phi\nabla^\nu\phi$ and
 $X\Box\phi$ modify the regularity condition on the horizon that determines the scalar charge of the black hole with respect to its mass and affects the regularity condition that determines the minimum black hole mass, whereas  $X^2$ leaves both conditions unaffected. All terms affect the scalar configuration however and a large positive coupling for $X^2$ can compromise the existence of black holes altogether. 
 
 Our two key findings are the following: (i) additional shift-symmetric interactions affect the minimum size of hair black holes, and hence the constraints one can derive from that ({\em e.g.}~\cite{2022arXiv220710692F, Charmousis:2021npl}), but their effect is rather moderate for dimensionless couplings (with respect to the scale of the black hole in geometric units) of order 1 or less. (ii) additional shift-symmetric interactions have an effect on the scaling of the charge per unit mass versus the mass of the black hole only for masses that are fairly close to the minimum mass. Hence, sufficiently large black holes in shift-symmetric theories will not carry a significant charge per unit mass, irrespective of the presence of additional shift-symmetric interactions [c.f. \cite{Saravani:2019xwx,Maselli:2020zgv,Maselli:2021men}]. 

\section{Acknowledgements}

T.P.S. acknowledges partial support from the STFC Consolidated Grants no. ST/T000732/1 and no. ST/V005596/1.
 
\appendix
\section{Equations of motion}
Varying action \eqref{eq:action} with respect to the scalar field yields the following scalar equation
\begin{align}\label{eq:scalar}
    \Box\phi=\,&-\alpha \mathcal{G}+2\gamma G^{\mu \nu } \nabla_{\nu }\nabla_{\mu }\phi+\kappa(\nabla\phi)^2\Box\phi+2\kappa \nabla^\mu\phi\nabla^\nu\phi\nabla_{\mu}\nabla_{\nu}\phi\\ \nonumber
    &-\sigma \nabla_{\mu }\,\Box\phi \nabla^{\mu}\phi -  \sigma\,(\Box\phi)^2 + \sigma\nabla^{\mu }\phi \,\Box\,\nabla_{\mu }\phi +\sigma(\nabla_{\mu}\nabla_{\nu }\phi)^2,
\end{align}
while varying with respect to the metric yields
\begin{equation}
\begin{split}
    \frac{1}{2}G_{\mu\nu}=
    &+\frac{1}{2}\nabla_\mu\phi\nabla_\nu\phi-\frac{1}{4}g_{\mu\nu}(\nabla\phi)^2\\
    &-\frac{\alpha}{2 g}g_{\mu(\rho}g_{\sigma)\nu}\epsilon^{\kappa\rho\alpha\beta}\epsilon^{\sigma\gamma\lambda\tau}\mathcal{R}_{\lambda\tau\alpha\beta}\nabla_{\gamma}\nabla_{\kappa}\phi\\
    &-\frac{\gamma}{2}  \mathcal{R} \nabla_{\mu }\phi \nabla_{\nu }\phi +  
    \gamma \nabla_{\nu }\nabla_{\mu }\phi \,\Box\phi - \gamma G_{\nu \rho } \nabla_{\mu }\phi \nabla^{\rho }
    \phi - \gamma G_{\mu \rho } \nabla_{\nu }\phi \nabla^{\rho 
    }\phi\\
    &+\frac{\gamma}{2} \mathcal{R}_{\mu \nu } (\nabla_{\mu }\phi)^2 - \gamma \nabla_{\rho }\nabla_{\nu }\phi \nabla^{\rho }\nabla_{\mu }\phi - \frac{\gamma}{2} g_{\mu \nu } (\Box\phi)^2 +  \frac{\gamma}{2} G_{\rho \sigma } g_{\mu \nu } \nabla^{\rho 
    }\phi \nabla^{\sigma }\phi\\
    &+\frac{\gamma}{2} g_{\mu \nu } \mathcal{R}_{\rho \sigma } \nabla^{\rho }\phi \nabla^{\sigma }\phi - \gamma 
    \mathcal{R}_{\mu \rho \nu \sigma } \nabla^{\rho }\phi \nabla^{\sigma 
    }\phi +  \frac{\gamma}{2} g_{\mu \nu } (\nabla_{\sigma }\nabla_{\rho }\phi)^2\\
    &+\frac{\sigma}{2} \nabla_{\mu}\phi \nabla_{\nu}\phi \;\Box\phi - \sigma\nabla_{\rho }\nabla_{(\mu}\phi\nabla_{\nu)}\phi \nabla^{\rho }\phi +  \frac{\sigma}{2} g_{\mu\nu} \nabla^{\rho }\phi \nabla_{\sigma }\nabla_{\rho }\phi \nabla^{\sigma }\phi\\
    &-\frac{\kappa}{2}(\nabla\phi)^2\nabla_\mu\phi\nabla_\nu\phi+\frac{\kappa}{8}g_{\mu\nu}(\nabla\phi)^2(\nabla\phi)^2.\\
\end{split}
\end{equation}

\section{Peturbative coefficients}
\label{sec:appendix_perturbative}
Here we present the coefficients appearing in the $\tilde{\alpha}$-expansion of the scalar charge and the ADM mass. In order to find the perturbative expressions we solve the perturbed field and scalar equations. To \textbf{first order}:
\begin{align}
    (tt):\quad &(r-2 m) B_1'+B_1=0,\\
    (rr):\quad&(2 m-r) A_1'+B_1=0,\\
    (\phi):\quad&r^5 (2 m-r) \phi _1''+2 r^4 (m-r) \phi_1'-48 m^2=0.
\end{align}
To \textbf{second order}:
\begin{align}
    \begin{split}
    (tt):\quad &m^2 r^{10} (r-2 m) B_2'+m^2 r^{10} B_2+2304 \gamma  m^6-384 \gamma  m^3 r^3\\
    &-52 m^2 r^6-8 m^3 r^5-16 m^4 r^4+736 m^5 r^3-2 m r^7+12 \gamma  r^6-r^8=0,
    \end{split}\\
    \begin{split}
    (rr):\quad &\left(4 m^2+2 m r+r^2\right) \left[8 m^3 \left(4\gamma +5 r^2\right)-16 m^2 r^3-4 r^3 \gamma+r^5\right]\\
    &+m^2 r^9 (2 m-r) A_2'+m^2 r^9 B_2=0,
    \end{split}\\
    \begin{split}
    (\phi):\quad &40 m^2 r^4 \sigma -672 m^3 r^3 \sigma -224 m^4 r^2 \sigma -2 m^2 r^{11} \phi _2'+2 m^3 r^{10} \phi _2'\\
    &-m^2 r^{12} \phi _2''+2 m^3 r^{11} \phi _2''-512 m^5 r \sigma +3456 m^6 \sigma +16 m r^5 \sigma +24 r^6 \sigma=0.
    \end{split}
\end{align}
Higher order equations are very lengthy but we have calculated them up to $\mathcal{O}(\tilde{\alpha}^5)$. We can then solve for the coefficients appearing in the expressions \eqref{eq:Qa}-\eqref{eq:Ma} for the charge and mass expansions:

\begin{align}
        \begin{split}
            Q_1=&\;\frac{2}{m},\;Q_3=-\frac{1}{60 m^5},\;Q_4=-\frac{689\sigma }{36960 m^9},\\[2mm]
            Q_5=&\;\frac{11051 \kappa }{720720 m^{11}}-\frac{268867 \gamma^2}{16336320 m^{13}}-\frac{84317 \gamma }{180180 m^{11}}-\frac{4609603 \sigma ^2}{130690560 m^{13}}-\frac{118549}{158400 m^9}.
        \end{split}\\[10mm]
        \begin{split}
            M_2=&\;\frac{49}{40 m^3},\; M_3=\;\frac{18107 \sigma }{73920 m^7},\\[2mm]
            M_4=&\;\frac{244007 \gamma }{360360 m^9}-\frac{635421 \gamma ^2}{10890880 m^{11}}+\frac{11838611 \sigma ^2}{87127040 m^{11}}+\frac{408253}{246400 m^7}+\frac{130309 \kappa }{1441440 m^9},\\[2mm]
            M_5=&\;\frac{995527207 \gamma  \sigma }{1241560320 m^{13}}-\frac{210006269 \gamma ^2 \sigma }{2595989760 m^{15}}+\frac{56711635 \kappa  \sigma }{496624128 m^{13}}+\frac{12276069473 \sigma ^3}{119189790720 m^{15}}\\[2mm]
            &\;+\frac{84509327587 \sigma }{50315865600 m^{11}}.
        \end{split}
\end{align}

In Eqs \eqref{eq:phi_pert}-\eqref{eq:GB_pert} we presented the general form of the expansions for the scalar field and the GB invariant up to the third order in $\tilde{\alpha}$, which were used in Fig.~\ref{fig:perturbative_treatment} Here we give the analytic expressions for the coefficients appearing in these expansions:
\begin{align}
\begin{split}
    \phi_0 = \,
    & 0\, ,  
\end{split}\\[5mm]
\begin{split}
    \phi_1= \, 
    & \frac{2 \alpha  \left(3 r^2+4 m^2+3 r\right)}{3 m r^3}\, ,  
\end{split}\\[5mm]
\begin{split}
    \phi_2 = \,
    & \frac{2 \sigma  \left(-224 m^5-84 m^4 r-24 m^3 r^2+84 m^2 r^3+42 m r^4+21 r^5\right)}{21 m^2 r^9}\, ,   
\end{split}\\[5mm]
\begin{split}
    \phi_3 = \,
    & \frac{1}{{30 m^4}}\bigg[\frac{18432 m^9 \sigma ^2}{r^{15}}+\frac{7680 m^8 \left(8 \gamma ^2+3 \sigma ^2\right)}{7 r^{14}}+\frac{15360 m^7 \left(8 \gamma ^2-\sigma ^2\right)}{13 r^{13}}+\\
    & +\frac{240 m (2 \gamma +\kappa )-952 m^3}{5 r^5}+\frac{48 m^2 \left(-10 \gamma ^2+58 m^4-5 m^2 (\gamma -\kappa )+15 \sigma ^2\right)}{r^8}\\
    & +\frac{320 m^6 \left(24 \gamma ^2+8 m^2 (7 \gamma
    -\kappa )-33 \sigma ^2\right)}{r^{12}}\\  
    & +\frac{96 m^4 \left(-8 \gamma ^2+24 m^2 (3 \gamma -\kappa )-5 \sigma ^2\right)}{r^{10}}+\frac{32 m^2 \left(15 (2 \gamma +\kappa )-82 m^2\right)}{3 r^6}+\\ 
    & +\frac{160 m \left(-6 \gamma ^2+35
    m^4+18 \kappa  m^2+12 \sigma ^2\right)}{7 r^7}+\frac{66 m^2}{r^4}+\frac{526 m}{3 r^3}-\frac{1}{2 m r}+\frac{73}{r^2}\\ 
    & +\frac{3840 m^5 \left(4 \gamma ^2+8 m^2 (4 \gamma -\kappa )-9 \sigma ^2\right)}{11 r^{11}}+\\ 
    & +\frac{160 m^3 \left(-72 \gamma ^2+424 m^4-24 m^2 (2 \gamma +\kappa )+99 \sigma ^2\right)}{9 r^9}\bigg] \, , 
\end{split}\\[5mm]
\begin{split}
    \gb_1 = \,    
    & \frac{48 m^2}{r^6} \, ,   
\end{split}\\[5mm]
\begin{split}
    \gb_2 = \,   
    & \frac{79872 \gamma  m^5}{r^{15}}+\frac{14336 \gamma m^4}{r^{14}}+\frac{6656 \gamma  m^3}{r^{13}}-\frac{12288 \gamma    m^2}{r^{12}}+\frac{53760 m^4}{r^{12}}\\   
    &-\frac{4096 m^3}{5 r^{11}}-\frac{448
        m^2}{r^{10}}+\frac{588}{5 m^2 r^6}-\frac{1408 \gamma  m}{r^{11}}+\frac{384 \gamma }{m r^9}-\frac{4608 m}{r^9}\\   
    &-\frac{64}{m r^7}-\frac{640 \gamma }{r^{10}}-\frac{32}{r^8} \, ,  
\end{split}\\[5mm]
\begin{split}
    \gb_3 = \,   
    &\frac{106496 \gamma  m^5 \sigma }{r^{19}}-\frac{5603328 \gamma  m^7 \sigma }{r^{21}}-\frac{98304 \gamma  m^6 \sigma }{r^{20}}
    +\frac{1892352 \gamma  m^4 \sigma }{r^{18}}\\  
    &+\frac{129024 \gamma  m^3 \sigma }{r^{17}}+\frac{28672 \gamma  m^2 \sigma }{r^{16}}-\frac{2174976 m^6 \sigma}{r^{18}}+\frac{3981312 m^5 \sigma }{11 r^{17}}\\   
    &+\frac{841728 m^4 \sigma }{5 r^{16}}+\frac{507904 m^3 \sigma }{r^{15}}-\frac{38016 m^2 \sigma }{r^{14}}+\frac{18107 \sigma}{770 m^6 r^6}-\frac{181248 \gamma  m \sigma }{r^{15}}\\  
    &+\frac{4608 \gamma  \sigma }{m^2 r^{12}}-\frac{4096 \gamma  \sigma }{m r^{13}}-\frac{122880 m \sigma }{7 r^{13}}-\frac{12288 \gamma  \sigma}{r^{14}}-\frac{27648 \sigma }{r^{12}}.    
\end{split}
\end{align}
\vspace{5mm}

\section*{References}
\bibliographystyle{iopart-num}
\bibliography{biblio.bib}

\end{document}